\documentclass[11pt,a4paper]{article}

\usepackage[utf8]{inputenc}
\usepackage[english]{babel}

\usepackage{amsmath,amsfonts,amssymb,mathtools,bm}

\usepackage[headheight=14pt,margin=2.5cm]{geometry}
\usepackage{fancyhdr}
\usepackage{graphicx}
\usepackage[dvipsnames]{xcolor}
\usepackage{tikz}
\usetikzlibrary{arrows.meta,positioning,calc,decorations.pathmorphing}
\usepackage{float}
\usepackage{subcaption}

\usepackage{booktabs,array,longtable}

\usepackage{listings}
\definecolor{codegreen}{rgb}{0,0.6,0}
\definecolor{codegray}{rgb}{0.5,0.5,0.5}
\definecolor{codepurple}{rgb}{0.58,0,0.82}
\definecolor{backcolour}{rgb}{0.95,0.95,0.92}
\lstdefinestyle{mystyle}{
  backgroundcolor=\color{backcolour},
  commentstyle=\color{codegreen},
  keywordstyle=\color{magenta},
  numberstyle=\tiny\color{codegray},
  stringstyle=\color{codepurple},
  basicstyle=\ttfamily\small,
  breaklines=true,
  captionpos=b,
  keepspaces=true,
  numbers=left,
  numbersep=5pt,
  showstringspaces=false,
  tabsize=2
}
\usepackage{tcolorbox}
\tcbuselibrary{skins,breakable}

\newtcolorbox{highlightbox}[1][]{%
  colback=teal!5!white,
  colframe=teal!75!black,
  fonttitle=\bfseries,
  boxrule=0.8pt, arc=4pt,
  left=2mm, right=2mm, top=1mm, bottom=1mm,
  #1}

\newtcolorbox{mathbox}[1][]{%
  colback=blue!3!white,
  colframe=blue!50!black,
  fonttitle=\bfseries,
  boxrule=0.6pt, arc=3pt,
  left=2mm, right=2mm, top=1mm, bottom=1mm,
  #1}

\usepackage{hyperref}
\hypersetup{
  colorlinks=true,
  linkcolor=blue!70!black,
  citecolor=OliveGreen,
  urlcolor=cyan!70!black,
  pdftitle={BioNetFlux — Descriptive Report},
  pdfauthor={S.~Bertoluzza}
}
\usepackage[numbers,sort&compress]{natbib}

\usepackage{titlesec}
\titleformat{\section}{\Large\bfseries\color{teal!80!black}}{\thesection}{1em}{}
\titleformat{\subsection}{\large\bfseries\color{teal!60!black}}{\thesubsection}{1em}{}

\newcommand{\code}[1]{\texttt{#1}}
\newcommand{\bionetflux}{\textsc{BioNetFlux}}

\newcommand{\ddt}{\frac{\partial}{\partial t}}
\newcommand{\ddx}{\frac{\partial}{\partial s}}
\newcommand{\norm}[1]{\lVert #1 \rVert}

\renewcommand{\hat}[1]{\widehat{#1}}

\begin{document}

\begin{titlepage}
  \centering

  \includegraphics[width=0.55\textwidth]{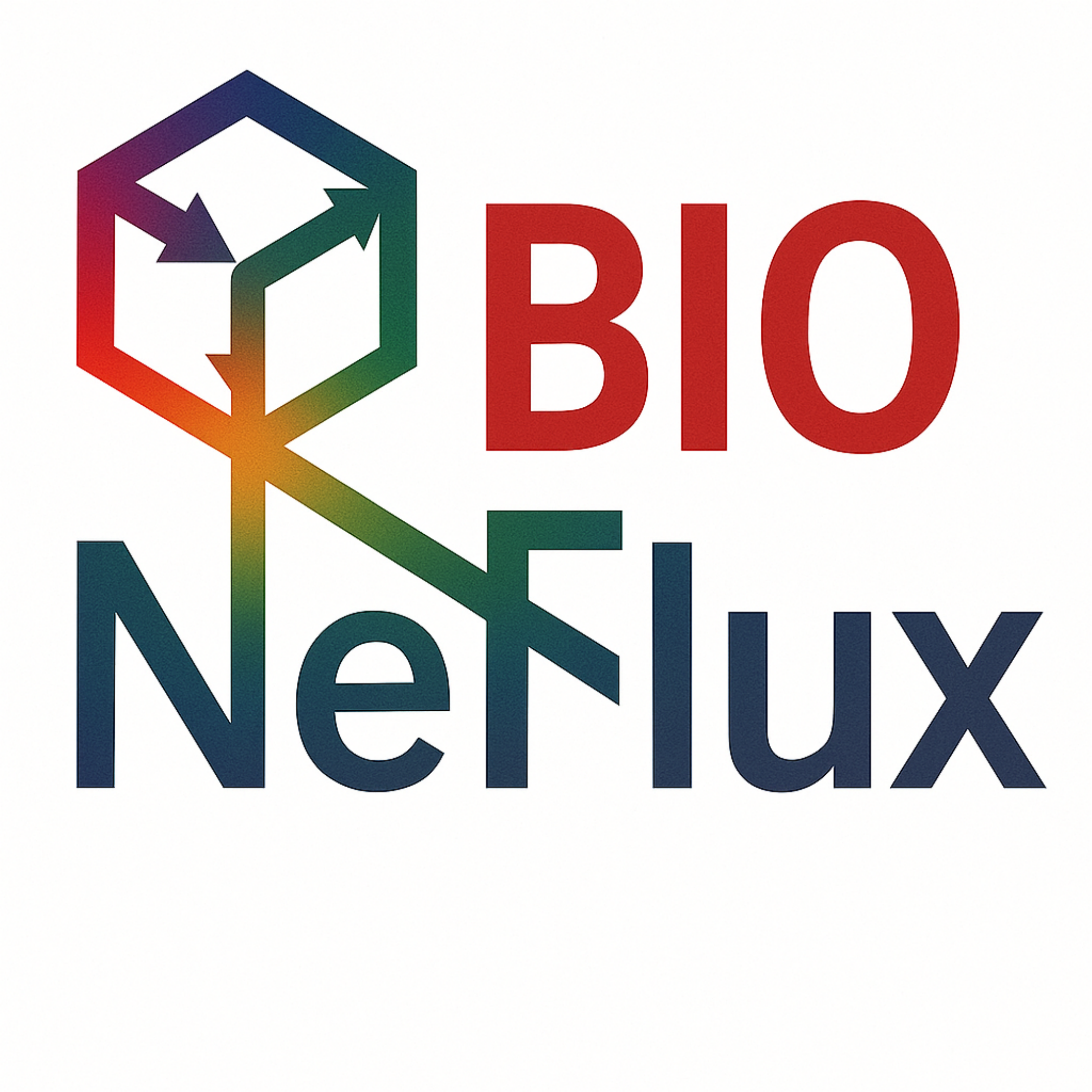}\\[1.2cm]

  {\Huge\bfseries BioNetFlux}\\[0.4cm]
  {\Large A Python Framework for Reaction--Diffusion--Chemotaxis\\
          Simulations on One-Dimensional Network Geometries}\\[2cm]

  {\large Silvia Bertoluzza}\\[0.3cm]
  {\normalsize Istituto di Matematica Applicata e Tecnologie Informatiche\\
               ``E.~Magenes'' --- CNR, Pavia, Italy}\\[2cm]

  {\large\today}

  \vfill

  {\footnotesize
   \textit{Hybridizable Discontinuous Galerkin methods \(\cdot\)
           Static condensation \(\cdot\)
           Adaptive time stepping}\\[0.2cm]
   \textit{Keller--Segel chemotaxis \(\cdot\)
           Organ-on-Chip transport \(\cdot\)
           Multi-arc maze geometries}}

  \vskip 2cm
  \includegraphics[width=1.1\textwidth]{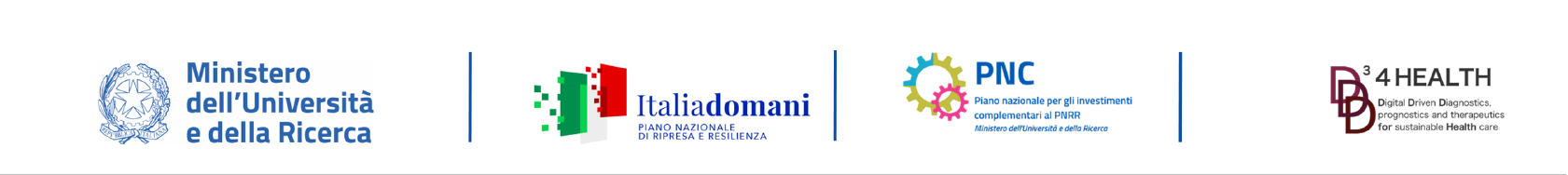}
\end{titlepage}

\tableofcontents
\clearpage

\section{Introduction}

\bionetflux{} is an open-source Python framework for the
numerical simulation of coupled systems of partial differential
equations (PDEs) on \emph{one-dimensional multi-arc networks}.
Its design targets biological transport phenomena\,---\,chemotaxis,
diffusion-reaction, and advection-diffusion\,---\,on graph-like
geometries that arise naturally in microfluidic organ-on-chip (OoC)
devices, vascular networks, and in-vitro cell-migration assays.
An AI language model (Claude) was used to assist in translating and extending an existing MATLAB implementation—originally written entirely by the author—into Python. The resulting Python code was reviewed, corrected, and fully validated by the author to ensure mathematical and numerical consistency with the MATLAB version.

\

The framework couples two methodological pillars:
\begin{enumerate}
  \item \textbf{Hybridizable Discontinuous Galerkin (HDG) spatial
        discretisation}~\cite{cockburn2009unified,cockburn2016static},
        which reduces the global system to unknowns living on the
        skeleton (in the 1D case, the nodes of the spatial decomposition and the network junctions) and
        recovers the interior solution by element-local back-solves;
  \item \textbf{Implicit  time integration  (backward-Euler)} with
        Newton--Raphson linearisation and optional adaptive
        time-step control, providing unconditional $A$-stability for
        the stiff chemotactic and reaction terms.
\end{enumerate}

Key capabilities include:
\begin{itemize}
  \item Arbitrary one-dimensional network topologies defined by
        point-and-line CSV files or programmatic builders;
  \item A modular problem layer with two built-in models\,---\,the
        \emph{Keller--Segel} (KS) chemotaxis system
        (2~coupled equations) and the \emph{Organ-on-Chip} (OoC) transport
        system (4~coupled equations)\,---\,and a template for user-defined
        extensions;
  \item Neumann, Dirichlet, Robin, trace-continuity, and
        Kedem--Katchalsky boundary/interface conditions;
  \item TOML-based configuration with symbolic function expressions
        resolved at load time;
  \item Rich visualisation: bird's-eye colour maps, flat-3D surface
        plots, domain-wise 2-D curves, and geometry diagrams.
\end{itemize}

The remainder of this report presents the mathematical models
(Section~\ref{sec:models}), the HDG discretisation
(Section~\ref{sec:hdg}), the software architecture
(Section~\ref{sec:architecture}), and a gallery of simulation
results obtained on a maze-type OoC geometry
(Section~\ref{sec:results}) mimicking one of the experiments in \cite{um2019immature}.

\section{Mathematical Models}\label{sec:models}

\subsection{Network Domain}

We consider a connected, planar graph
$\mathcal{G} = (\mathcal{V},\mathcal{E})$
whose edges $e_j\in\mathcal{E}$, $j=1,\dots,N_{\mathrm{dom}}$,
are one-dimensional segments of length~$L_j$.
Each edge is independently parametrised by a local coordinate
$s\in[x_0,x_0+L_j]$.  Default value for $x_0$ is $x_0=0$. Vertices in $\mathcal{V}$ are classified as:
\begin{itemize}
  \item \textbf{Junction points (J)}: shared endpoints of two or more
        edges; trace-continuity or membrane Kedem-Kachalski conditions, as well as Kirchhoff flux-balance conditions
        are imposed here.
  \item \textbf{T-junction points (T)}: endpoints of one edge that
        touch the interior of another edge; conditions can be imposed similar to the previous ones.
  \item \textbf{Boundary points (B)}: endpoints that lie on the
        exterior boundary of the overall domain; Dirichlet, Neumann, or Robin conditions
        are imposed.
\end{itemize}

\subsection{Keller--Segel Chemotaxis System (2 Equations)}

The Keller--Segel model~\cite{keller1970initiation,perthame2007transport}
describes the coupled dynamics of a cell density~$u(s,t)\geq 0
$ and a chemoattractant concentration~$\varphi(s,t)\geq 0$:
\begin{mathbox}[title={Keller--Segel system}]
\begin{align}
  \ddt u &= \nu\,\frac{\partial^2 u}{\partial s^2}
            - \ddx\!\Bigl[\chi(\varphi)\,u\,\ddx\varphi\Bigr]
            + f_u,
  \label{eq:ks-u}\\[4pt]
  \ddt \varphi &= \mu\,\frac{\partial^2 \varphi}{\partial s^2}
                  + b\,u - a\,\varphi + f_\varphi,
  \label{eq:ks-phi}
\end{align}
\end{mathbox}
\noindent
where
\begin{itemize}
  \item $\nu,\mu>0$ are the diffusion coefficients of cells and
        chemoattractant, respectively;
  \item $\chi(\varphi)$ is the chemotactic sensitivity function
        (see below);
  \item $b>0$ is the rate of chemoattractant production by cells;
  \item $a\geq 0$ is the chemoattractant natural decay rate;
  \item $f_u,f_\varphi$ are external source/sink terms.
\end{itemize}

\paragraph{Chemotaxis sensitivity.}
\bionetflux{} offers different options for the definition of the chemotaxis sensitivity function $\chi$. Among other things, it  implements the \emph{receptor-saturation} model:
\begin{equation}\label{eq:chi}
  \chi(\varphi) = \frac{k_1}{(k_2+\varphi)^2}\,,
  \qquad
  \chi'(\varphi) = \frac{-2\,k_1}{(k_2+\varphi)^3}\,,
\end{equation}
where $k_1$ is the cellular drift velocity and $k_2$ the receptor
dissociation constant.  This saturating form prevents the finite-time
blow-up that occurs with constant sensitivity~$\chi\equiv\chi_0$.

\subsection{Organ-on-Chip Transport System (4 Equations)}

The OoC model extends the chemotaxis framework to describe
immune-cell migration, tumour-cell dynamics, and two signalling
molecules on a microfluidic network.  The unknowns are
$(u,\omega,v,\varphi)$, with the following physical roles:
\begin{center}
\begin{tabular}{clcl}
  \toprule
  Symbol & Meaning & Diffusivity & Eq.~index \\
  \midrule
  $u$        & immune-cell density        & $\nu$       & 0 \\
  $\omega$   & chemoattractant~$\omega$   & $\epsilon$  & 1 \\
  $v$        & tumour-cell density        & $\sigma$    & 2 \\
  $\varphi$  & chemoattractant~$\varphi$  & $\mu$       & 3 \\
  \bottomrule
\end{tabular}
\end{center}

\begin{mathbox}[title={Organ-on-Chip system}]
\begin{align}
  \ddt u &= \nu\,\frac{\partial^2}{\partial s^2} u
            - \frac{\partial}{\partial s} \bigl[\chi(\varphi)\,u\,\frac{\partial}{\partial s}\varphi_s\bigr]
            + f_u,
  \label{eq:ooc-u}\\[3pt]
  \ddt\omega &= \epsilon\,\frac{\partial^2}{\partial s^2}\omega
                - c\,\omega + d\,u + f_\omega,
  \label{eq:ooc-omega}\\[3pt]
  \ddt v &= \sigma\,\frac{\partial^2}{\partial s^2}v
            - \lambda(\omega)\,v + f_v,
  \label{eq:ooc-v}\\[3pt]
  \ddt\varphi &= \mu\,\frac{\partial^2}{\partial s^2}\varphi
                 - a\,\varphi + b\,u + f_\varphi,
  \label{eq:ooc-phi}
\end{align}
\end{mathbox}

\noindent
The coupling structure is as follows:
\begin{itemize}
  \item \textbf{Immune cells}~$u$ undergo diffusion with coefficient
        $\nu$ and \emph{chemotaxis} driven by the gradient of
        $\varphi$, with sensitivity $\chi(\varphi)$ as
        in~\eqref{eq:chi}.
  \item \textbf{Chemoattractant~$\omega$} diffuses with coefficient
 $\epsilon$,
        decays at rate~$c$, and is produced by immune cells at
        rate~$d$.
  \item \textbf{Tumour cells}~$v$ diffuse with coefficient
  $\sigma$ 
        and are suppressed by $\omega$ via the
        function $\lambda(\omega)=m_1/(m_2+\omega)$. 
  \item \textbf{Chemoattractant~$\varphi$} diffuses with coefficient
  $\mu$,
        decays at rate~$a$, and is produced by immune cells at
        rate~$b$.
\end{itemize}
The system captures a biologically relevant feedback loop:
immune cells produce signal molecules that guide their own
migration (chemotaxis) while simultaneously suppressing tumour
growth. Stationary tumor cells can be emulated by setting $\sigma\ll 1$ as well as $\lambda(\omega) = 0$.

\subsection{Boundary and Interface Conditions}\label{sec:bc}

At the boundary and at the junctions of the network the user may assign, independently
for each equation, the following condition: at the boundary:
\begin{description}
  \item[Neumann] $D\,\partial_s u\,n = g(t)$\,---\,prescribed
        flux (homogeneous by default).
  \item[Dirichlet] $u = g(t)$\,---\,prescribed trace.
  \item[Robin] $\alpha\,u + \beta\,D\,\partial_s u\,n = g(t)$.
\end{description}
At junctions and T-junctions flux balance is imposed together with one of the following coupling conditions:
\begin{description}
  \item[Trace continuity] $u^{(j_1)}=u^{(j_2)}$ at a junction
        shared by edges $j_1,j_2$, together with a Kirchhoff-type
        flux balance.
  \item[Kedem--Katchalsky]
        $J_s=\omega_{\mathrm{KK}}(u^{(j_1)}-u^{(j_2)})$\,---\,%
        membrane-permeability condition modelling selective barriers
        in OoC devices.
\end{description}

\section{HDG Spatial Discretisation}\label{sec:hdg}

Hybridizable Discontinuous Galerkin methods
\cite{cockburn2009unified,cockburn2016static}
introduce a \emph{numerical trace}
$\hat u$ on the mesh skeleton and reformulate the PDE as a first-order
system.  The key advantage is that, after \emph{static condensation},
the only globally coupled unknowns are the traces at element
interfaces and network junctions\,---\,the interior (bulk) solution
is recovered by cheap, embarrassingly parallel element-local
back-solves.

\subsection{Local Problem on Element~$K$}

Consider a generic scalar equation
$\partial_t u = D\,\partial_{ss}u + R(u,\dots)$ on an element
$K=[s_L,s_R]$ of length $h$.  Introduce the auxiliary flux variable
$q = -D\,\partial_s u$ and seek piecewise-polynomial approximations
$(u_h,q_h)\in\mathbb{P}_p(K)\times\mathbb{P}_{p_q}(K)$ satisfying
\begin{align}
  \int_K \frac{u_h^{n+1}-u_h^n}{\Delta t}\,w\,\mathrm{d}s
  - \int_K q_h^{n+1}\,w'\,\mathrm{d}s
  + [q_h^{n+1} w \nu]\Big|_{\partial K}
  + \tau\bigl[u_h^{n+1}-\hat u^{n+1}\bigr]\Big|_{\partial K}
  &= \int_K R^{n+1}\,w\,\mathrm{d}s,
  \label{eq:hdg-u}\\[4pt]
  \int_K D^{-1}q_h\,r\,\mathrm{d}s
  + \int_K u_h\,r'\,\mathrm{d}s
  - \hat u\,r\Big|_{\partial K}
  &= 0,
  \label{eq:hdg-q}
\end{align}
for all test functions $w\in\mathbb{P}_p(K)$,
$r\in\mathbb{P}_{p_q}(K)$, where $\tau>0$ is the HDG
stabilisation parameter and $\nu$ is the unit vector pointing outwards of $K$, so that
\[
[w ]\Big|_{\partial K} = w(s_R) + w(s_L), \qquad [w \nu ]\Big|_{\partial K} = w(s_R) - w(s_L).
\]

\subsection{Coupling and Boundary Conditions}
Coupling between elements is achieved by defining the numerical trace $\hat u$ at the element interfaces as  single valued at the nodes of the decomposition. The local equations \eqref{eq:hdg-u}-\eqref{eq:hdg-q} are complemented by coupling equations imposing the flux balance conditions. For $K^- = [s_L,\widehat s]$ and $K^+ = [\widehat s, s_R]$, with $s_L < \hat s < s_R$, we set
\[
q^- = q^{n+1}_h |_{K^-},\qquad q^+ = q^{n+1}_h|_{K^+}, \qquad u^- = u^{n+1}_h |_{K^-},\qquad u^+ = u^{n+1}_h|_{K^+}
\]
and the flux balance is imposed as
\begin{equation}\label{eq:hdg-flux}
q^+(\widehat s) - q^-(\widehat s) - \tau [ u^+(\widehat s) + u^-(\widehat s) - 2\widehat u^{n+1}(\widehat s)] = 0.
\end{equation}
Junction and boundary conditions are implemented by introducing, for all arcs, Lagrange multipliers at all concerned nodes. The flux balance for the given domain is modified by adding the corresponding 
multiplier and, depending on whether it is a junction or boundary node, one or two additional equations are added to the system to enforce the desired condition. 
More precisely, for a boundary condition of the form $\alpha u + \beta q = g$ at a 
boundary node, the equation 
\[
\alpha \hat u + \beta \lambda = g,
\]
is added to the system, where $\lambda$ is the Lagrange multiplier associated to 
the  domain containing the boundary node. 
For continuity at a junction node $\hat s$ shared by $2$ domains, an  equations of the form
\[\hat u^{(1)}(\hat s) - \hat u^{(2)}(\hat s) = 0,\]
as well as the coupling flux balance equation
\[
\lambda^{(1)}(\hat s) + \lambda^{(2)}(\hat s) = 0.
\]

\subsection{Static Condensation}

Equations~\eqref{eq:hdg-u},\eqref{eq:hdg-q},\eqref{eq:hdg-flux} can be written in
matrix form on each element as
\begin{equation}\label{eq:sc-block}
  \begin{pmatrix} A_{ii} & A_{ib} \\ A_{bi} & A_{bb}\end{pmatrix}
  \begin{pmatrix} \bm u_i \\ \hat{\bm u}\end{pmatrix}
  =
  \begin{pmatrix} \bm f_i \\ \bm f_b\end{pmatrix},
\end{equation}
where $\bm u_i$ collects the interior DOFs and $\hat{\bm u}$ the
trace DOFs,
and where the first line encompasses equations \eqref{eq:hdg-u}--\eqref{eq:hdg-q}, while the second line stands for equation \eqref{eq:hdg-flux}.
  Eliminating the bulk unknown $\bm u_i$ requires solving:
\[
  \widetilde A\,\hat{\bm u} = \widetilde{\bm f},
  \qquad
  \widetilde A = A_{bb}-A_{bi}A_{ii}^{-1}A_{ib},
  \qquad
  \widetilde{\bm f} = \bm f_b - A_{bi}A_{ii}^{-1}\bm f_i.
\]
The condensed contributions from all elements are assembled into a
\emph{global} system whose unknowns are the trace values at all
inter-element faces and network junctions, plus Lagrange multipliers
that enforce flux balance.  After solving this reduced system, the
bulk solution is recovered on each element independently. Remark that the matrix $A_{ii}$ is block diagonal with small blocks that can be inverted exactly independently of each other thus making the static condensation procedure embarassingly parallel.

\subsection{Polynomial Orders}

At present
\bionetflux{} implements order $1$ HDG, but it offers the possibility of using  different polynomial orders for the flux variable, thus implementing two different versions of the lowest order HDG method. More precisely, for the two model considered,
depending on the coupling structure of each equation:
\begin{itemize}
  \item \textbf{Keller--Segel}: equation~0 ($u$): $\mathbb{P}_0$
        flux; equation~1 ($\varphi$): $\mathbb{P}_1$ flux.
  \item \textbf{Organ-on-Chip}: equation~0 ($u$): $\mathbb{P}_0$
        flux; equations~1--3 ($\omega,v,\varphi$): $\mathbb{P}_1$
        flux.
\end{itemize}
These choices balance accuracy and computational cost: the
chemotaxis equation for~$u$ couples nonlinearly to~$\varphi$ and
requires only a low-order flux representation, while the remaining
diffusion-reaction equations benefit from a richer flux space.
Higher order methods can be added to the code by writing corresponding static condensation modules.

\subsection{Elementary Matrices}

All element matrices are pre-computed on the reference element
$\hat K=[0,1]$ using SymPy for exact symbolic integration, then
scaled analytically to the physical element length~$h$.  The cached
matrices include the mass matrix~$M$, its inverse~$M^{-1}$, the
differentiation matrix~$D$, trace-evaluation operators
$\widetilde N,\hat N$, boundary matrices $G_b,M_b,T$, an
averaging operator~$A_v$, as well as a quadrature matrix~$Q$ with
associated nodes. 

\subsection{Newton--Raphson Solver}

After backward-Euler time discretisation the nonlinear algebraic
system
$\bm R(\bm U^{n+1})=\bm 0$
is solved by Newton's method:
\begin{equation}
  J(\bm U^k)\,\delta\bm U = -\bm R(\bm U^k),
  \qquad
  \bm U^{k+1}=\bm U^k+\delta\bm U.
\end{equation}
Convergence is declared when
$\norm{\bm R}_2<\varepsilon_{\text{abs}}$.
Two globalization strategies are available:
\emph{damped Newton} ($\bm U^{k+1}=\bm U^k+\alpha\,\delta\bm U$
with fixed~$\alpha\in(0,1]$) and
\emph{line-search Newton} (backtracking until
$\norm{\bm R(\bm U^{k+1})}<\norm{\bm R(\bm U^k)}$).

\subsection{Adaptive Time Stepping}

\bionetflux{} implements an adaptive strategy that adjusts $\Delta t$
based on the Newton iteration count~$n_{\text{it}}$ at each step:
\[
  \Delta t_{\text{new}} =
  \begin{cases}
    1.2\,\Delta t & \text{if } n_{\text{it}} \leq 8\quad
                     \text{(easy step)},\\
    \Delta t       & \text{if } 9\leq n_{\text{it}}\leq 14\quad
                     \text{(moderate)},\\
    0.8\,\Delta t & \text{if } n_{\text{it}} > 14\quad
                     \text{(hard step)},\\
    0.5\,\Delta t & \text{if Newton fails; retry the step}.
  \end{cases}
\]
This heuristic keeps the time step near the edge of efficient Newton
convergence without requiring error estimators.

\section{Software Architecture}\label{sec:architecture}

\bionetflux{} is distributed as an installable Python package
(\texttt{pip install -e .}) requiring Python~$\geq$\,3.11,
NumPy, Matplotlib, SymPy, and optionally \texttt{tomli} (for
Python~$<$\,3.11).  The source tree is organised into
sub-packages that mirror the mathematical workflow:

\begin{highlightbox}[title=Package layout]
\small
\begin{verbatim}
src/bionetflux/
+-- core/             Problem, Discretization, Constraints,
|                     BulkData, DomainData, BulkDataManager,
|                     StaticCondensation, FluxJump,
|                     GlobalAssembly
+-- geometry/         DomainGeometry, DomainInfo, ConnectionInfo,
|                     builder functions, maze CSV parser
+-- problems/         ks_problem, ooc_problem, config managers
+-- time_integration/ NewtonSolver, TimeStepper,
|                     AdaptiveTimeStepper
+-- utils/            ElementaryMatrices, ConfigManager
+-- visualization/    LeanMatplotlibPlotter
+-- analysis/         ErrorEvaluator, MinimalErrorEvaluator
+-- setup_solver.py   SolverSetup orchestrator, quick_setup()
\end{verbatim}
\end{highlightbox}

\paragraph{Data flow.}
\begin{enumerate}
  \item The user selects a \emph{problem module}
        (\code{ks\_problem} or \code{ooc\_problem}) and a
        TOML configuration file.
  \item \code{quick\_setup()} imports the module, calls
        \code{create\_global\_framework()}, builds or receives a
        \code{DomainGeometry}, and initialises a
        \code{SolverSetup} object that lazily creates elementary
        matrices, static-condensation operators, the global assembler,
        and the bulk-data manager.
  \item A \code{TimeStepper} (or \code{AdaptiveTimeStepper}) advances
        the solution, calling at each time step:
        bulk data update $\to$ static condensation $\to$ global
        assembly of residual and Jacobian $\to$ Newton solve $\to$
        bulk recovery.
  \item At user-specified intervals the \code{LeanMatplotlibPlotter}
        produces bird's-eye views, 3-D surface plots, or domain-wise
        2-D curves.
\end{enumerate}

\paragraph{Geometry from CSV.}
Network topologies can be defined  by two CSV files
(\code{points.csv}, \code{lines.csv}).  Each point carries a
tag whose first letter classifies it as \textbf{J}~(junction),
\textbf{T}~(T-junction), or \textbf{B}~(boundary).  The parser
automatically generates all interior and exterior connections,
and the optional \code{length} scaling parameter maps unit-square
layouts to physical dimensions.

\paragraph{Configuration.}
Physical parameters, time-stepping controls, initial and boundary
conditions, and per-domain overrides are all specified in a single
TOML file.  Symbolic expressions such as
\code{"2.5 + 0.0*s"} or \code{"sin(2*pi*s)"} are resolved to
Python callables at load time via a three-stage cascade
(literal~$\to$ symbolic~$\to$ built-in library).

\section{Simulation Results}\label{sec:results}

We illustrate the capabilities of \bionetflux{} with a
representative simulation of the four-equation OoC system
\eqref{eq:ooc-u}--\eqref{eq:ooc-phi}
on a \emph{maze geometry} consisting of~29 segments with multiple
junctions, T-junctions, and boundary nodes, built from a pair of
CSV files (\code{maze\_3\_data}).  All coordinates are scaled by
a factor $\ell=50$, yielding segment lengths in the range
$[50,300]$ length units. This simulation mimicks one of the experiments described in \cite{um2019immature}.

\paragraph{Parameters.}
The simulation uses the parameters listed in
Table~\ref{tab:params}.  The chemotaxis sensitivity follows
``receptor saturation''
$\chi(\varphi)={k_1}/{(k_2+\varphi)^2}$
with $k_1=3.9\times 10^{-9}$ and $k_2=5\times 10^{-6}$.
Tumour suppression is deactivated ($m_1=0$).
Initial conditions are zero everywhere except for a localised
tumour mass $v=2.5$ on domain~17 and an immune-cell injection
$u=1$ on domain~28.

\begin{table}[H]
  \centering
  \caption{Physical parameters for the OoC maze simulation.}
  \label{tab:params}
  \begin{tabular}{@{}llrl@{}}
    \toprule
    Group & Symbol & Value & Description \\
    \midrule
    Diffusion
      & $\nu$ (immune cells)       & 200    & $D_u$ \\
      & $\epsilon$ (chemoattr.~$\omega$) & 900 & $D_\omega$ \\
      & $\sigma$ (tumour cells)    & $10^{-9}$ & $D_v$\,(quasi-immobile) \\
      & $\mu$ (chemoattr.~$\varphi$) & 900 & $D_\varphi$ \\[3pt]
    Reaction
      & $a$ ($\varphi$ decay)      & $10^{-4}$ & $\beta_\varphi$ \\
      & $c$ ($\omega$ decay)       & $10^{-4}$ & $\beta_\omega$ \\[3pt]
    Coupling
      & $b$ ($\varphi$ production) & 0.2   & $\alpha_\varphi$ \\
      & $d$ ($\omega$ production)  & 0.1   & $\alpha_\omega$ \\[3pt]
    Time
      & $T$                        & 640  & final time \\
      & $ \Delta t_0$               & 64   & initial time step \\[3pt]
    Discretisation
      & $h$                        & 15    & target mesh size \\
      & $\tau$                     & $(0.5,0.5,0.5,0.5)$ & HDG stabilisation \\
    \bottomrule
  \end{tabular}
\end{table}

\subsection{Maze Geometry}

\begin{figure}
  \centering
  \includegraphics[width=0.70\textwidth]{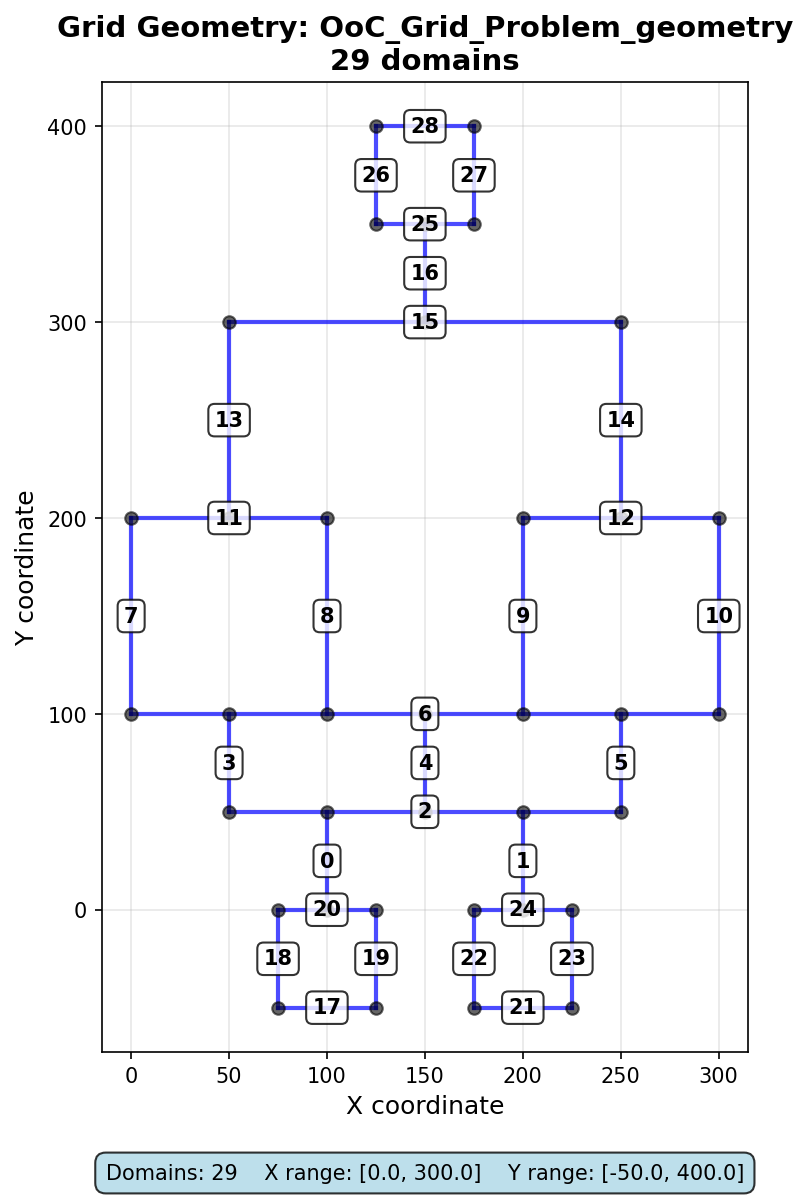}
  \caption{Maze geometry (\code{maze\_3\_data}, scaled by
           $\ell=50$) with domain indices.  Blue numbers label
           segments; red dots mark junctions (J), T-junctions (T),
           and boundary points (B).}
  \label{fig:geometry}
\end{figure}

Figure~\ref{fig:geometry} shows the network geometry with domain
indices.  The maze features two ``inlet'' boxes at the bottom
(domains~17--24), a central area with multiple junctions and
T-junctions, an ``outlet'' box at the top (domains~25--28), and
a rich interior connectivity requiring careful treatment of
multi-point constraints.

\subsection{Solution Snapshots --- Bird's-Eye Views}

\begin{figure}
  \centering
  \begin{subfigure}[t]{0.32\textwidth}
    \includegraphics[width=\textwidth]{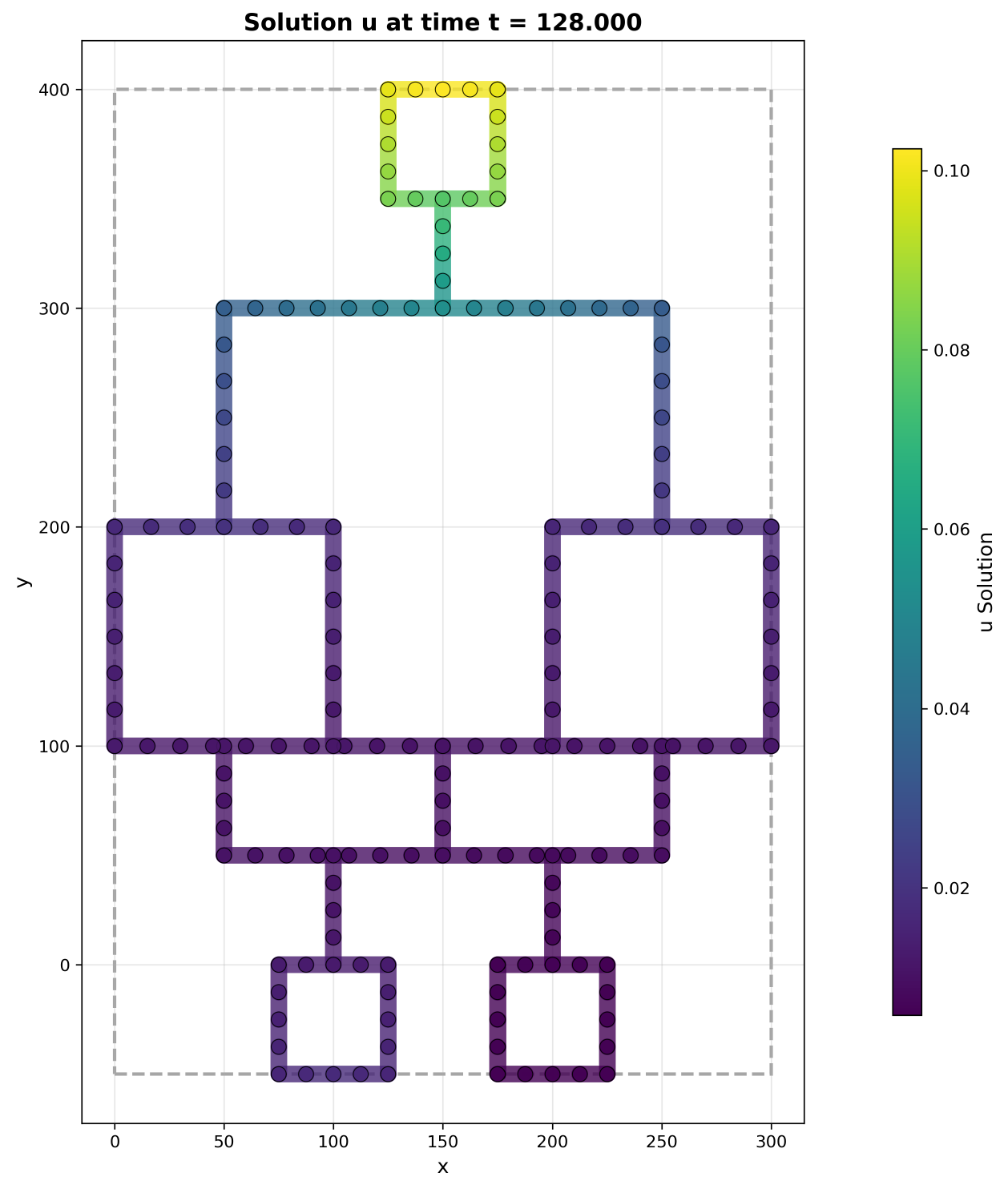}
    \caption{$t=0.25\,T$}
  \end{subfigure}\hfill
  \begin{subfigure}[t]{0.32\textwidth}
    \includegraphics[width=\textwidth]{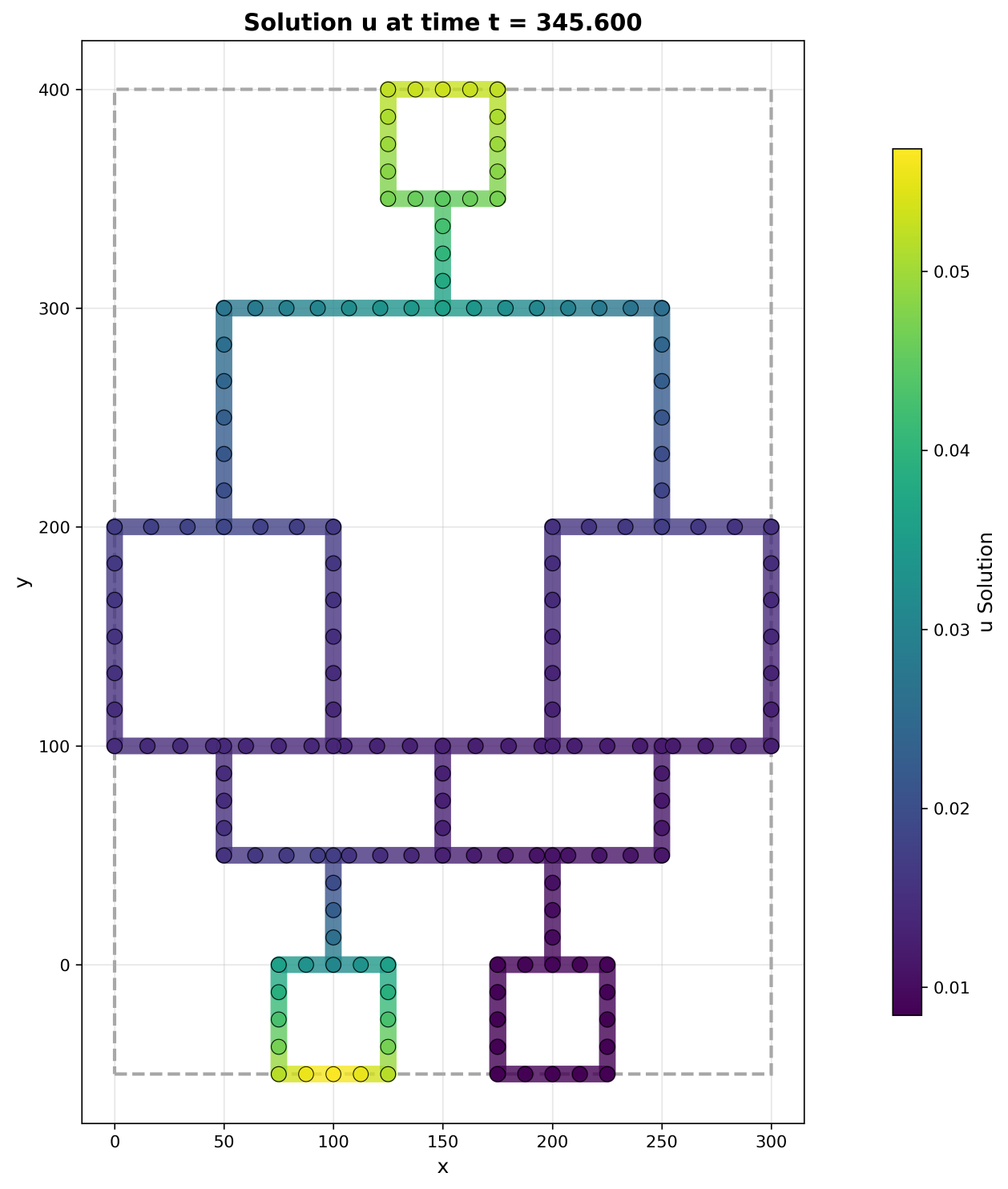}
    \caption{$t=0.50\,T$}
  \end{subfigure}\hfill
  \begin{subfigure}[t]{0.32\textwidth}
    \includegraphics[width=\textwidth]{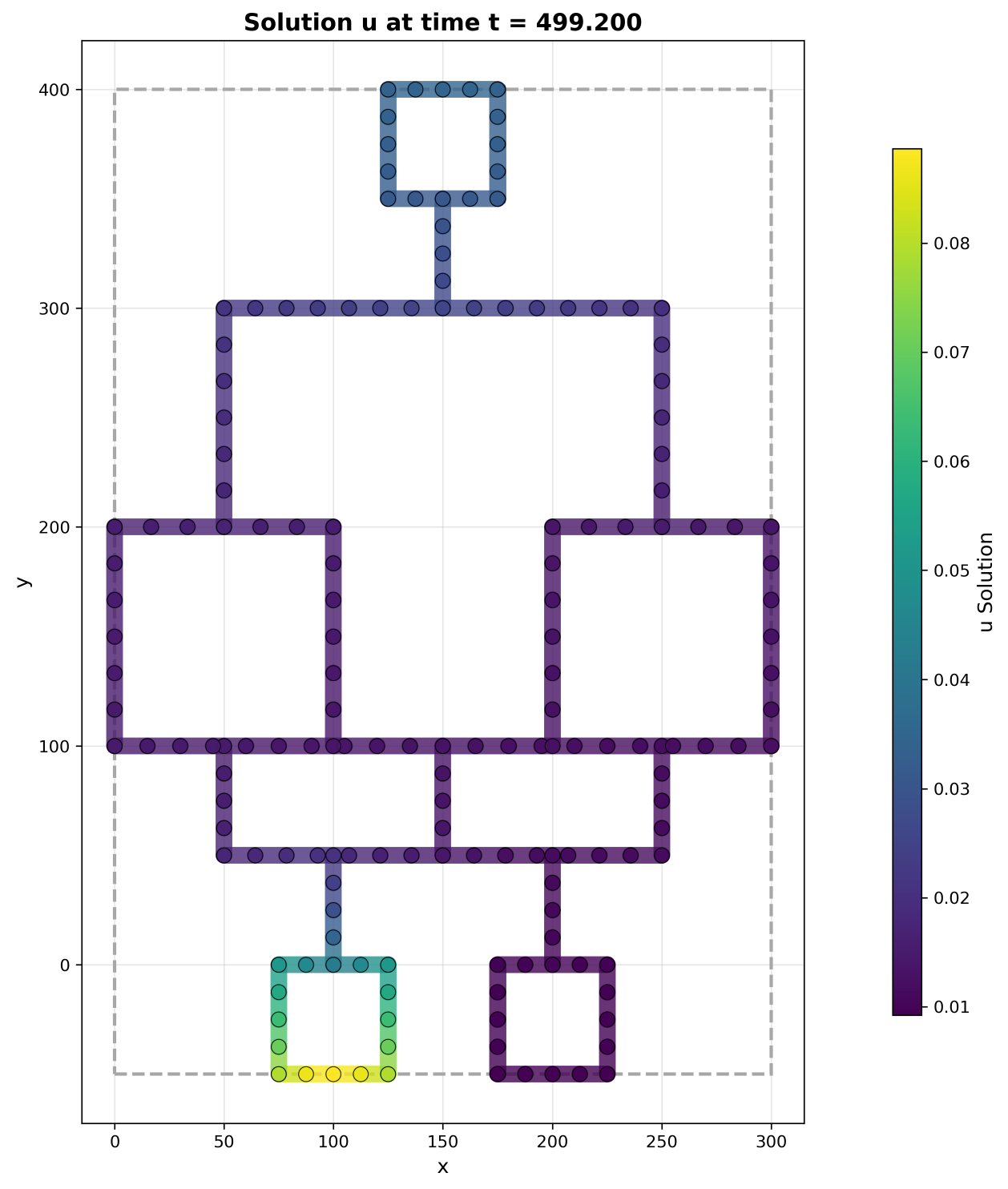}
    \caption{$t=0.75\,T$}
  \end{subfigure}
  \caption{Immune-cell density~$u$ (equation~0) at three time
           snapshots.  The initial injection at domain~28
           diffuses and migrates toward the chemoattractant
           sources via chemotaxis.}
  \label{fig:birdview-u}
\end{figure}

\begin{figure}
  \centering
  \begin{subfigure}[t]{0.32\textwidth}
    \includegraphics[width=\textwidth]{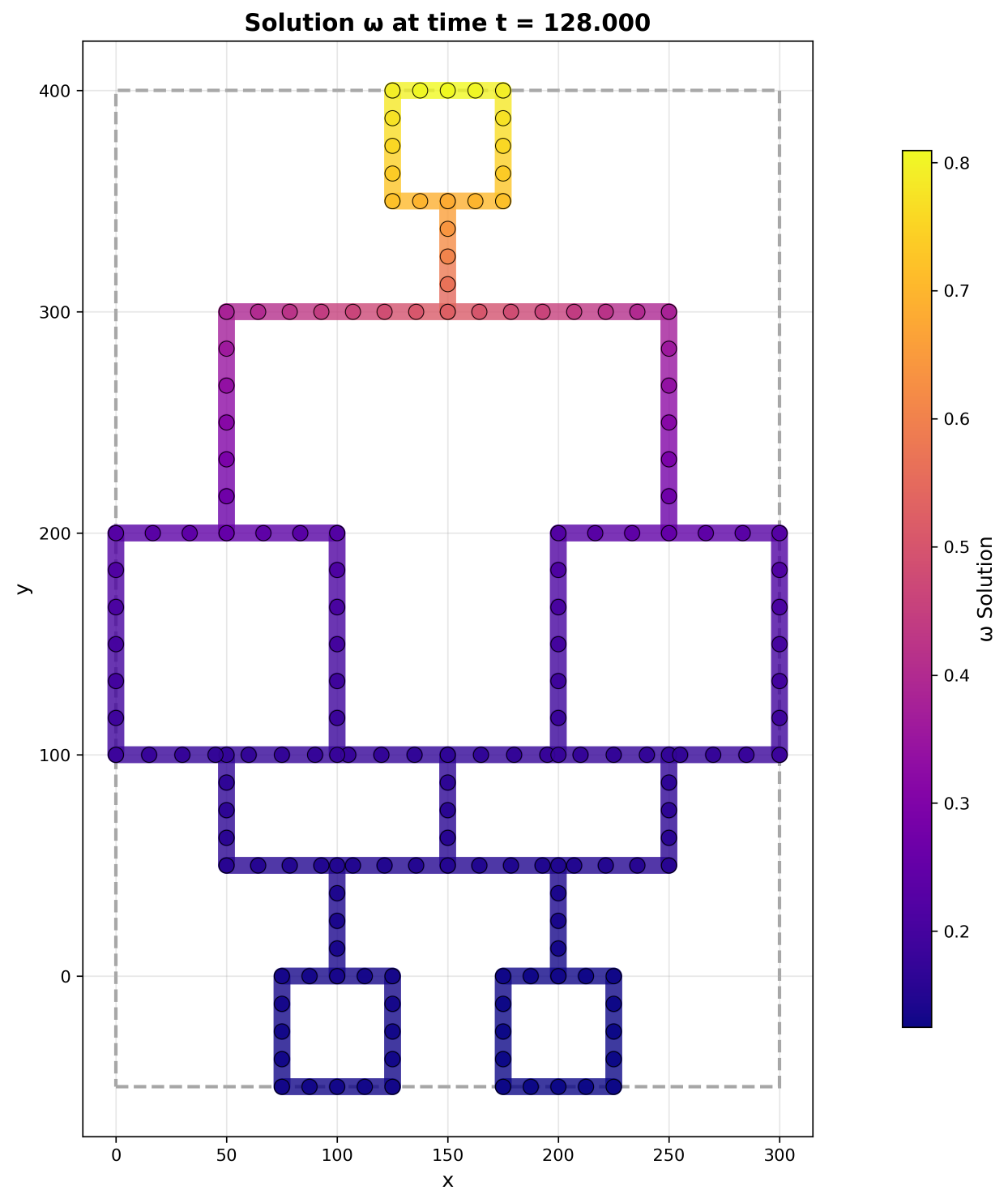}
    \caption{$t=0.25\,T$}
  \end{subfigure}\hfill
  \begin{subfigure}[t]{0.32\textwidth}
    \includegraphics[width=\textwidth]{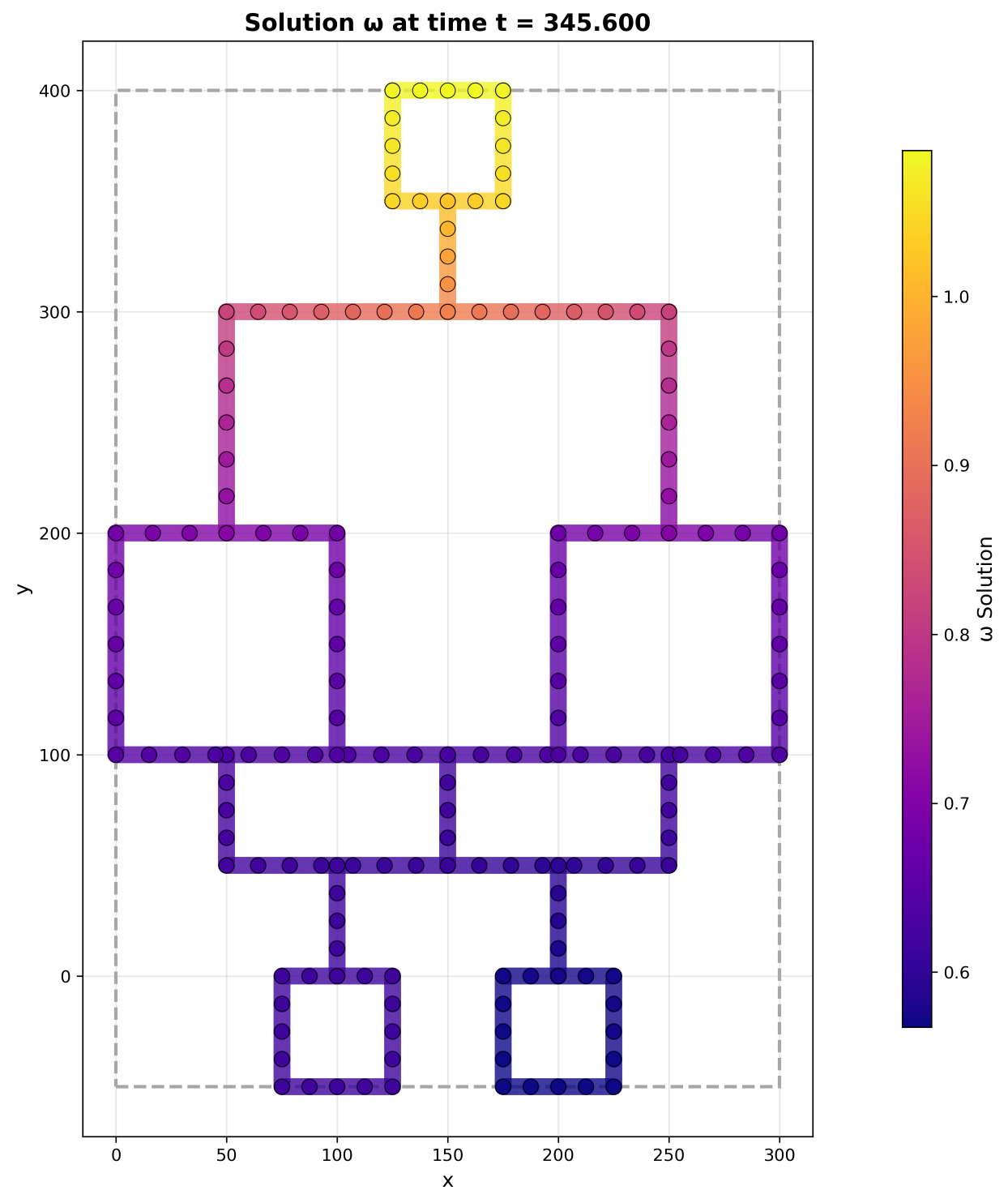}
    \caption{$t=0.50\,T$}
  \end{subfigure}\hfill
  \begin{subfigure}[t]{0.32\textwidth}
    \includegraphics[width=\textwidth]{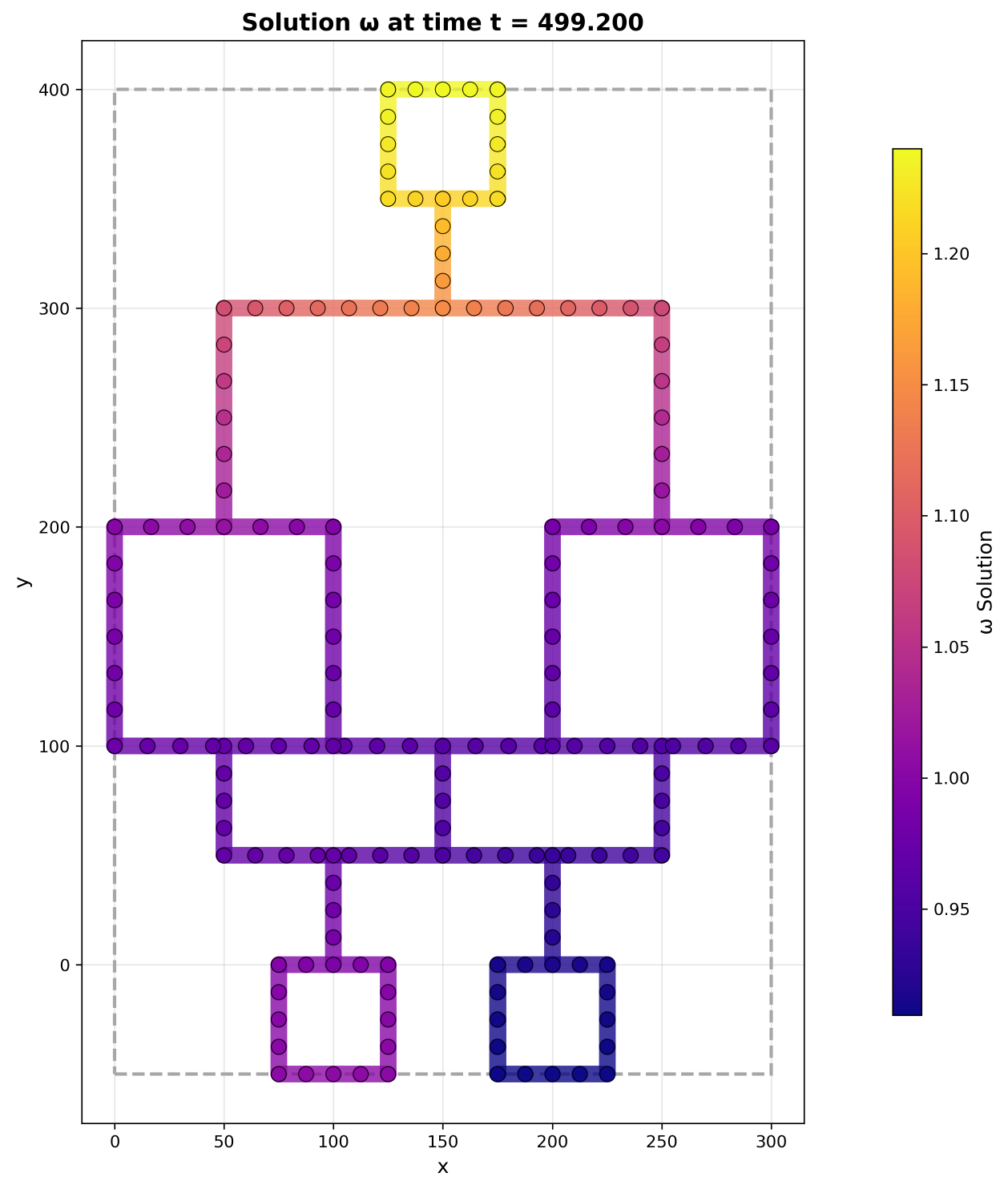}
    \caption{$t=0.75\,T$}
  \end{subfigure}
  \caption{Chemoattractant~$\omega$ (equation~1).  Produced
           by immune cells, $\omega$~diffuses rapidly
           ($\epsilon=900$) and decays slowly ($c=10^{-4}$).}
  \label{fig:birdview-omega}
\end{figure}

\begin{figure}
  \centering
  \begin{subfigure}[t]{0.32\textwidth}
    \includegraphics[width=\textwidth]{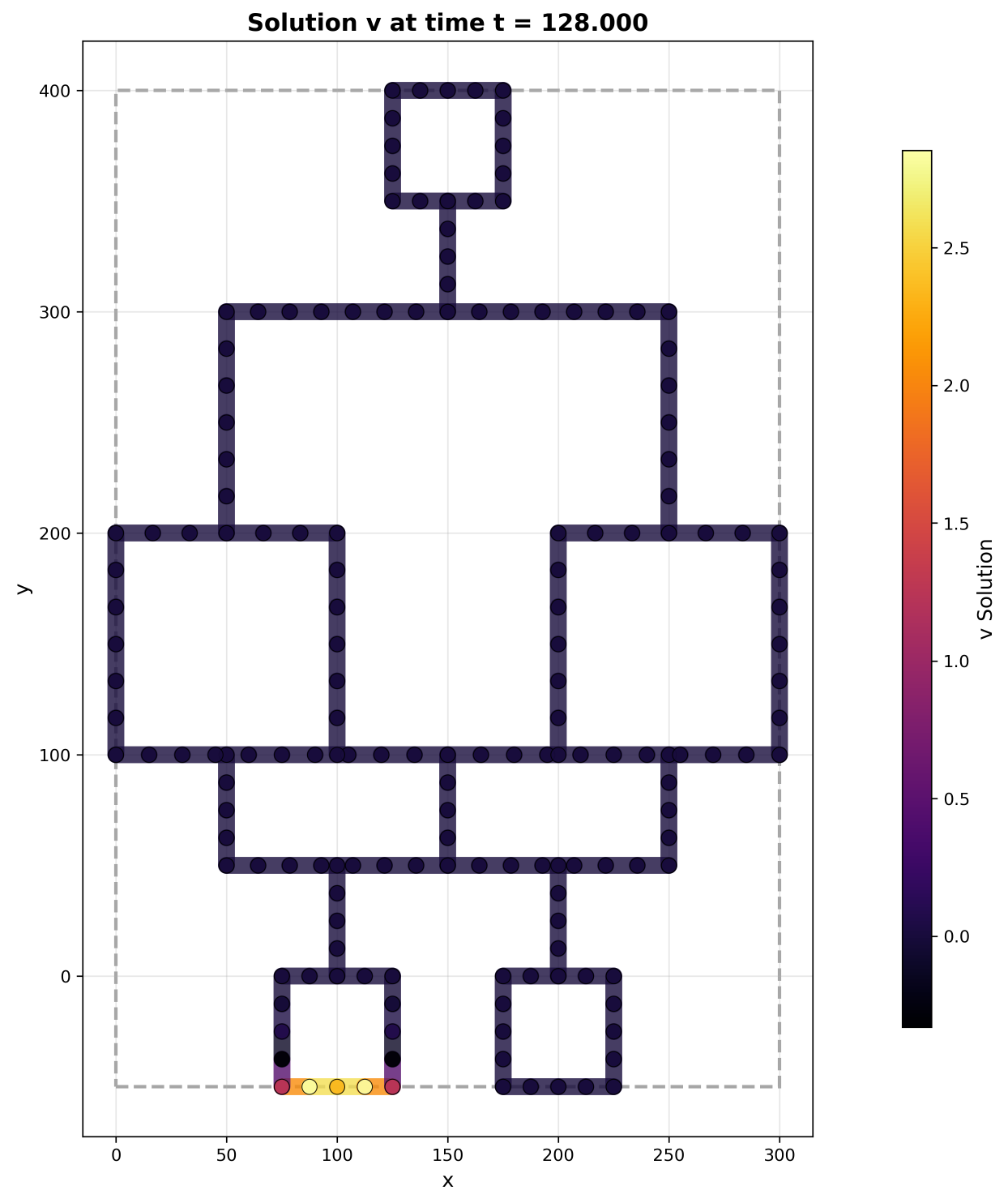}
    \caption{$t=0.25\,T$}
  \end{subfigure}\hfill
  \begin{subfigure}[t]{0.32\textwidth}
    \includegraphics[width=\textwidth]{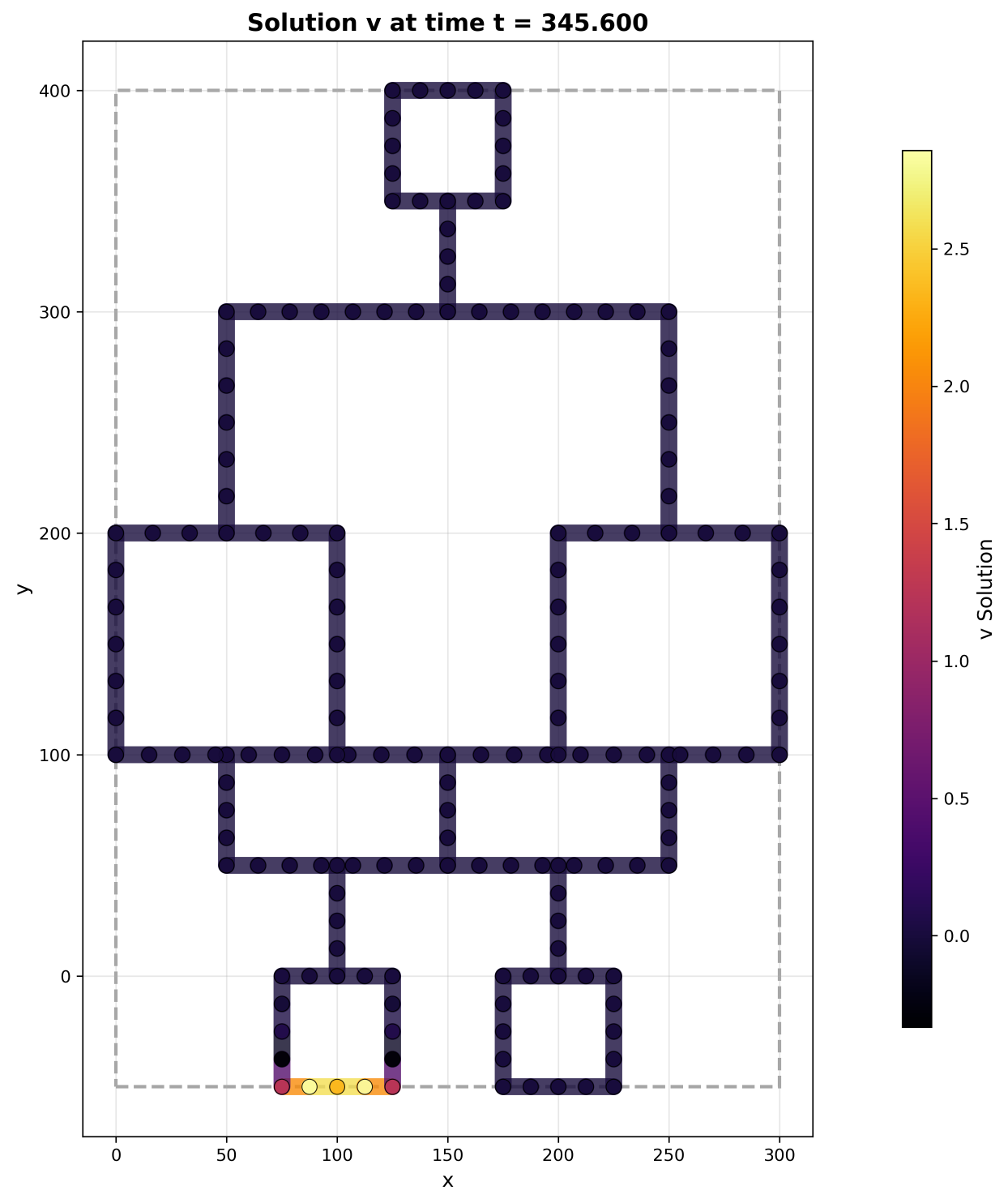}
    \caption{$t=0.50\,T$}
  \end{subfigure}\hfill
  \begin{subfigure}[t]{0.32\textwidth}
    \includegraphics[width=\textwidth]{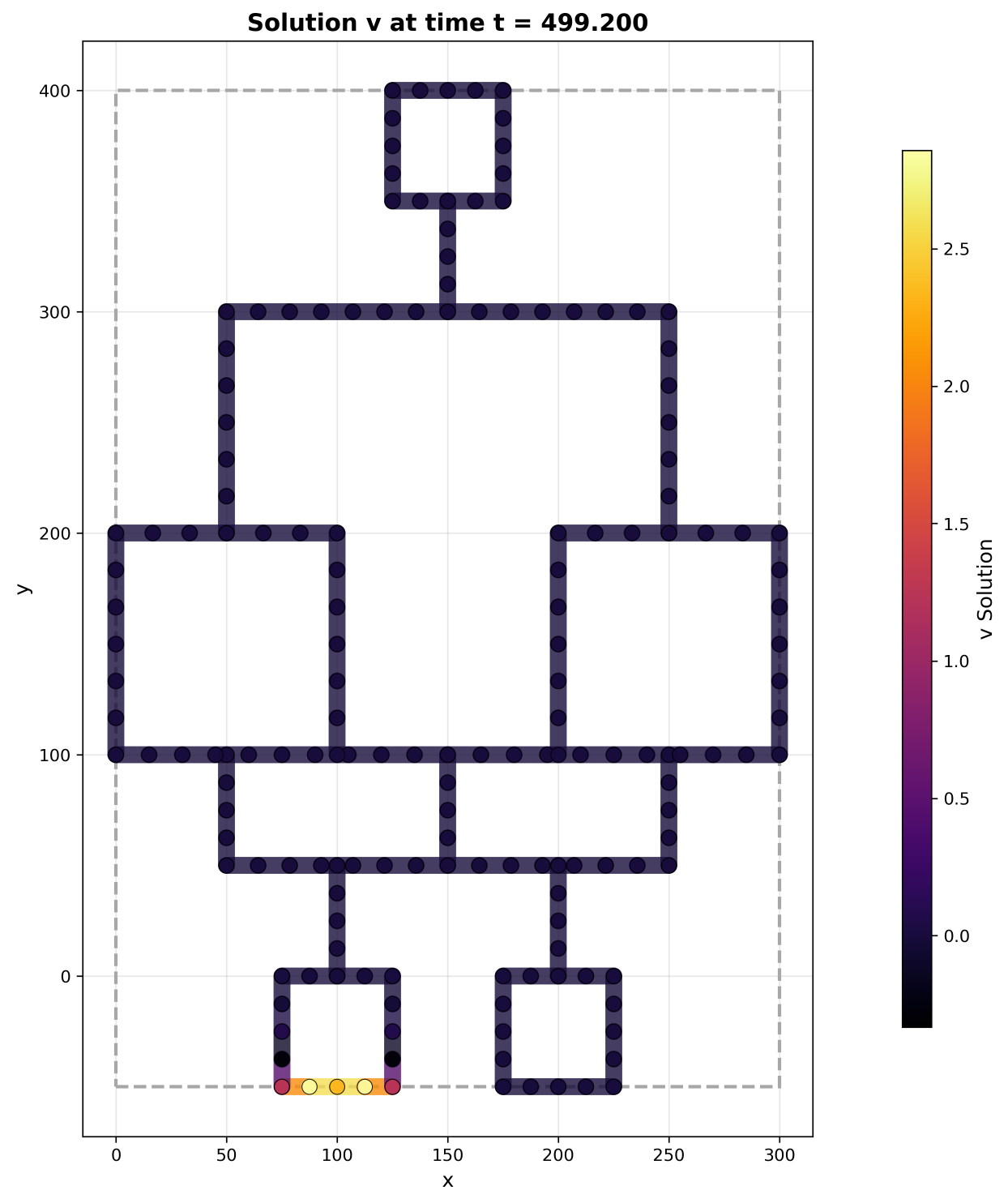}
    \caption{$t=0.75\,T$}
  \end{subfigure}
  \caption{Tumour-cell density~$v$ (equation~2).  With near-zero
           diffusivity~$\sigma=10^{-9}$, for this experiment, the tumour mass remains
           localised on domain~17 throughout the simulation.}
  \label{fig:birdview-v}
\end{figure}

\begin{figure}
  \centering
  \begin{subfigure}[t]{0.32\textwidth}
    \includegraphics[width=\textwidth]{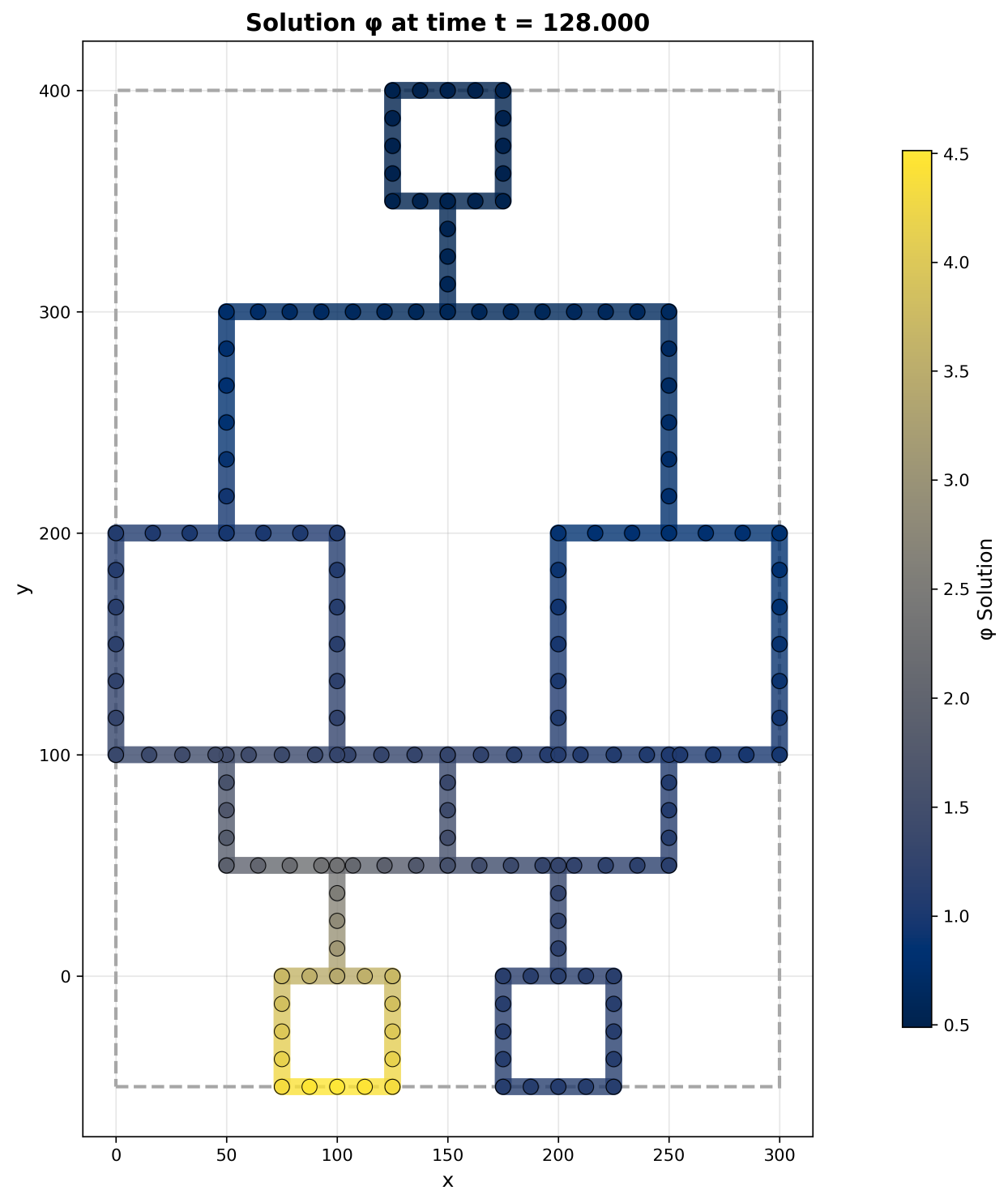}
    \caption{$t=0.25\,T$}
  \end{subfigure}\hfill
  \begin{subfigure}[t]{0.32\textwidth}
    \includegraphics[width=\textwidth]{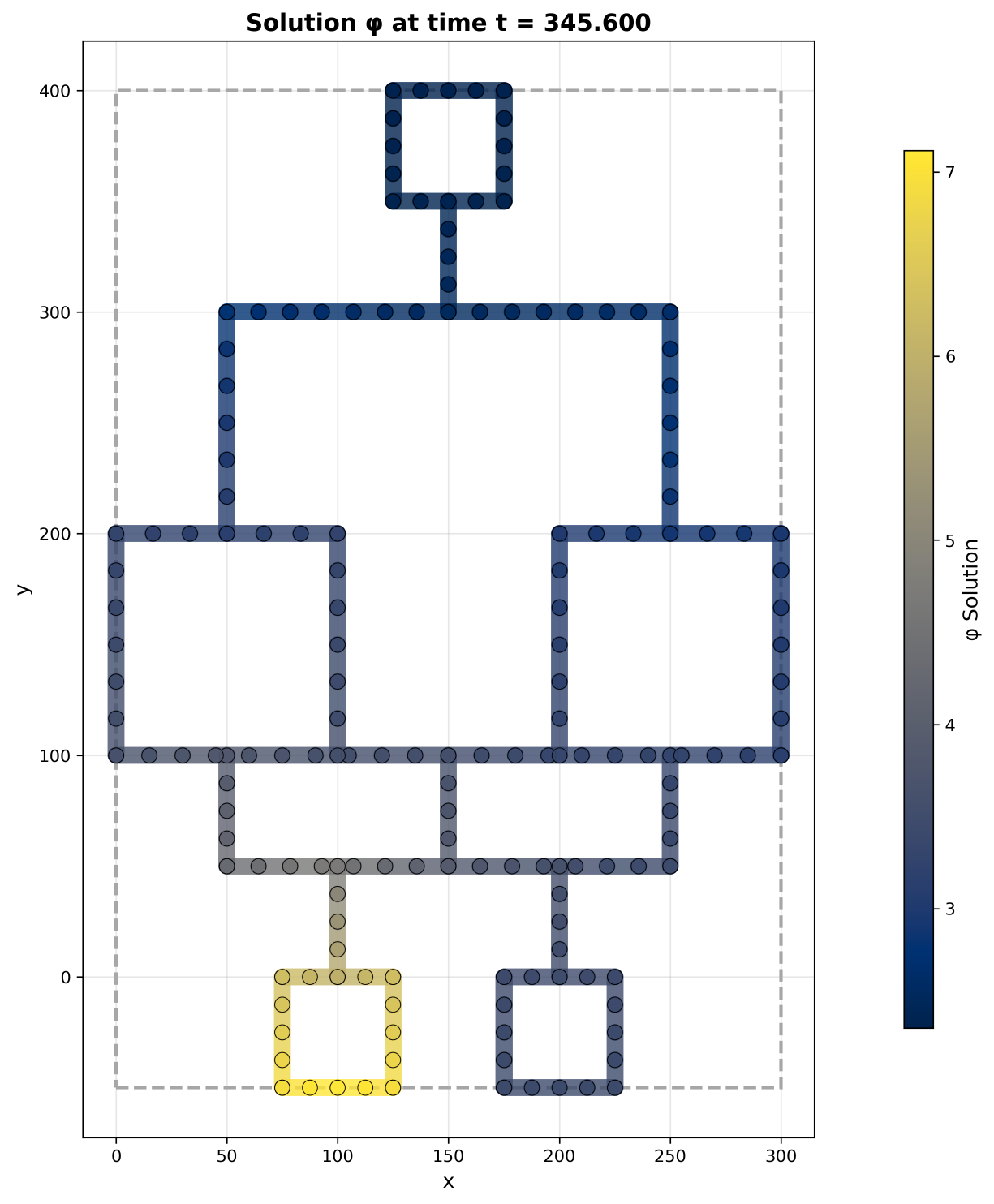}
    \caption{$t=0.50\,T$}
  \end{subfigure}\hfill
  \begin{subfigure}[t]{0.32\textwidth}
    \includegraphics[width=\textwidth]{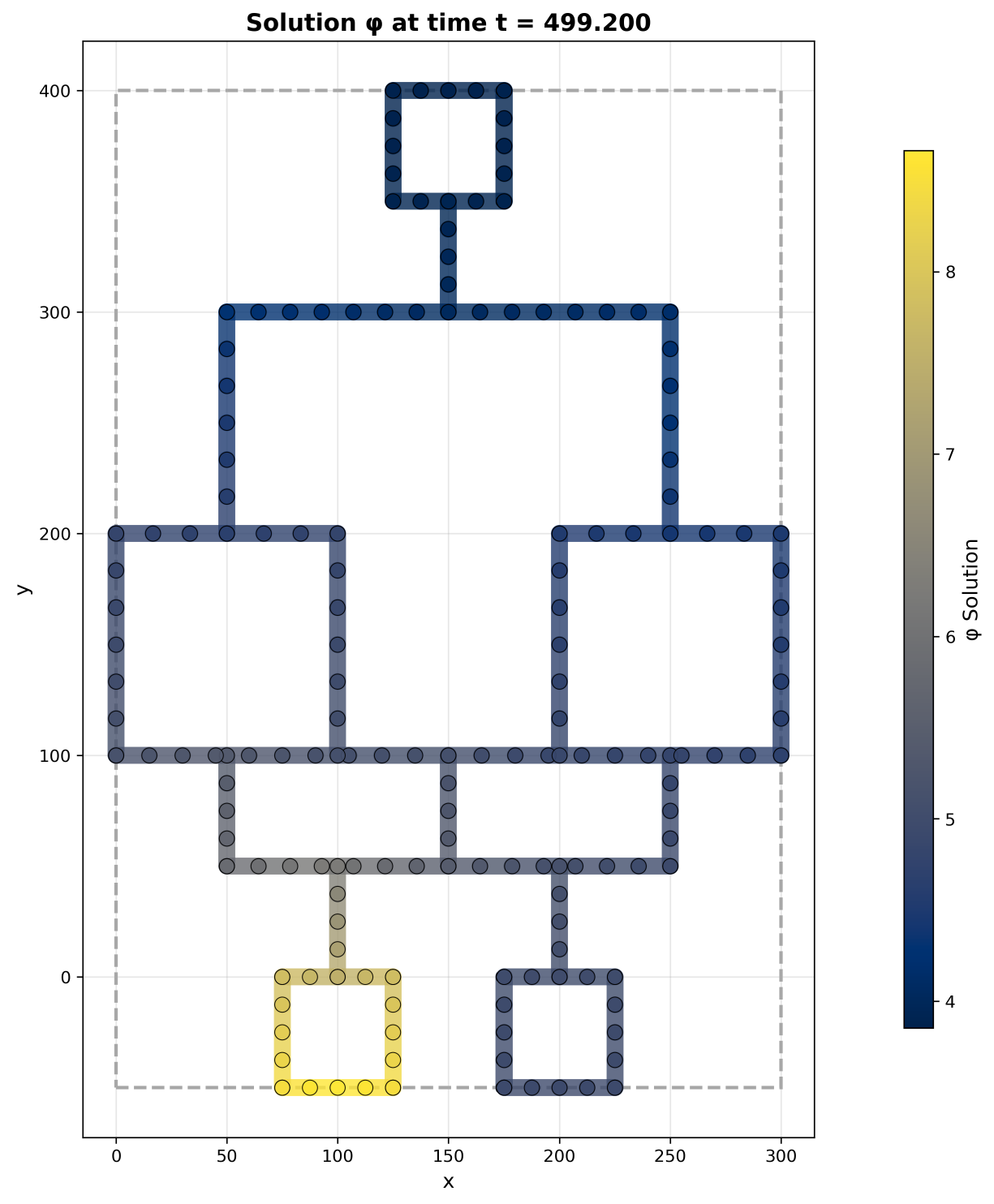}
    \caption{$t=0.75\,T$}
  \end{subfigure}
  \caption{Chemoattractant~$\varphi$ (equation~3).  Produced
           by immune cells at rate~$b=0.2$, $\varphi$ guides the
           chemotactic response of~$u$.}
  \label{fig:birdview-phi}
\end{figure}

Figures~\ref{fig:birdview-u}--\ref{fig:birdview-phi} present
colour-coded bird's-eye views of the four solution
components at 25\%, 50\%, and 75\% of the final time~$T$.

\subsection{Final State}

\begin{figure}
  \centering
  \begin{subfigure}[t]{0.48\textwidth}
    \includegraphics[width=\textwidth]{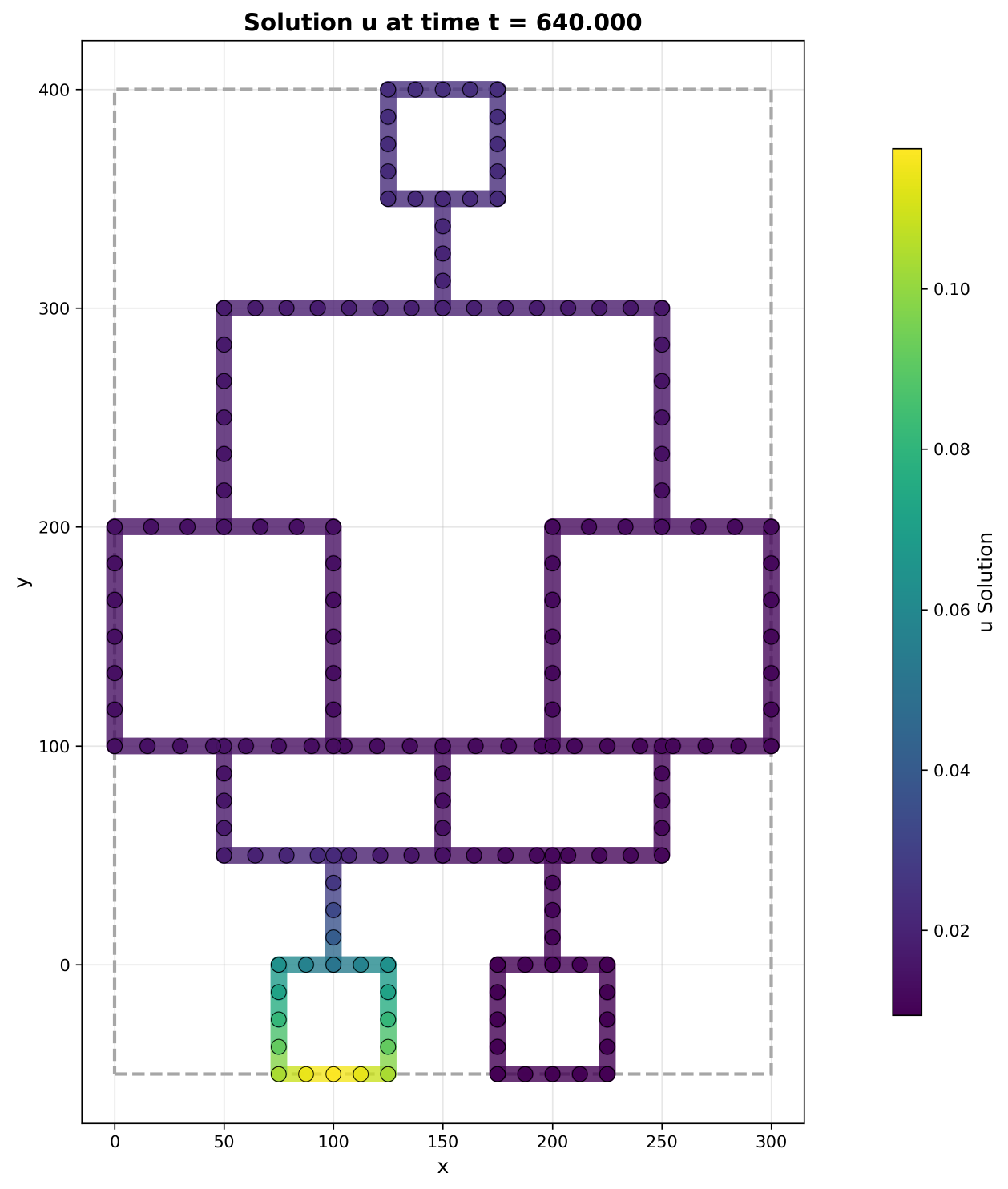}
    \caption{Immune cells~$u$}
  \end{subfigure}\hfill
  \begin{subfigure}[t]{0.48\textwidth}
    \includegraphics[width=\textwidth]{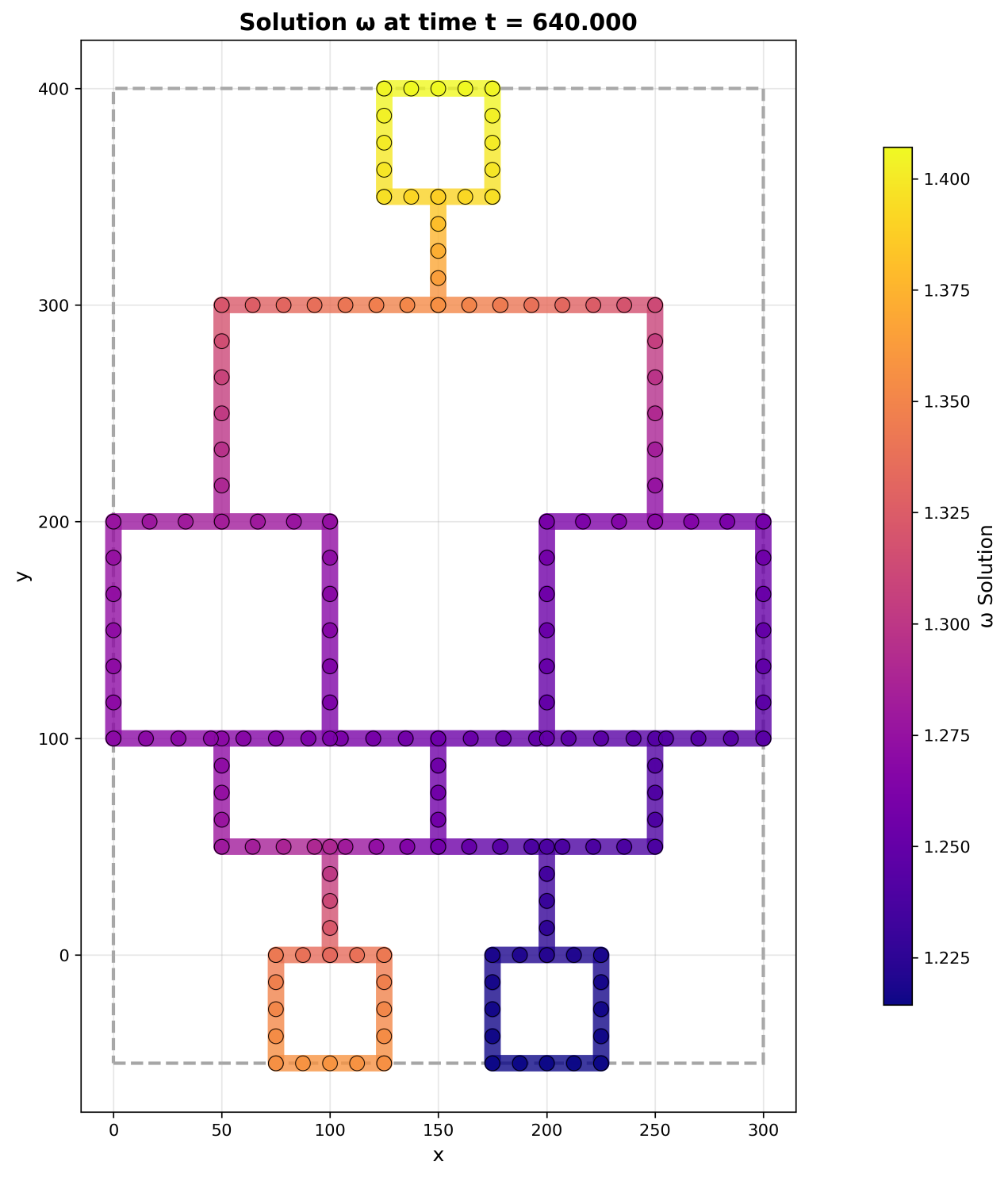}
    \caption{Chemoattractant~$\omega$}
  \end{subfigure}\\[6pt]
  \begin{subfigure}[t]{0.48\textwidth}
    \includegraphics[width=\textwidth]{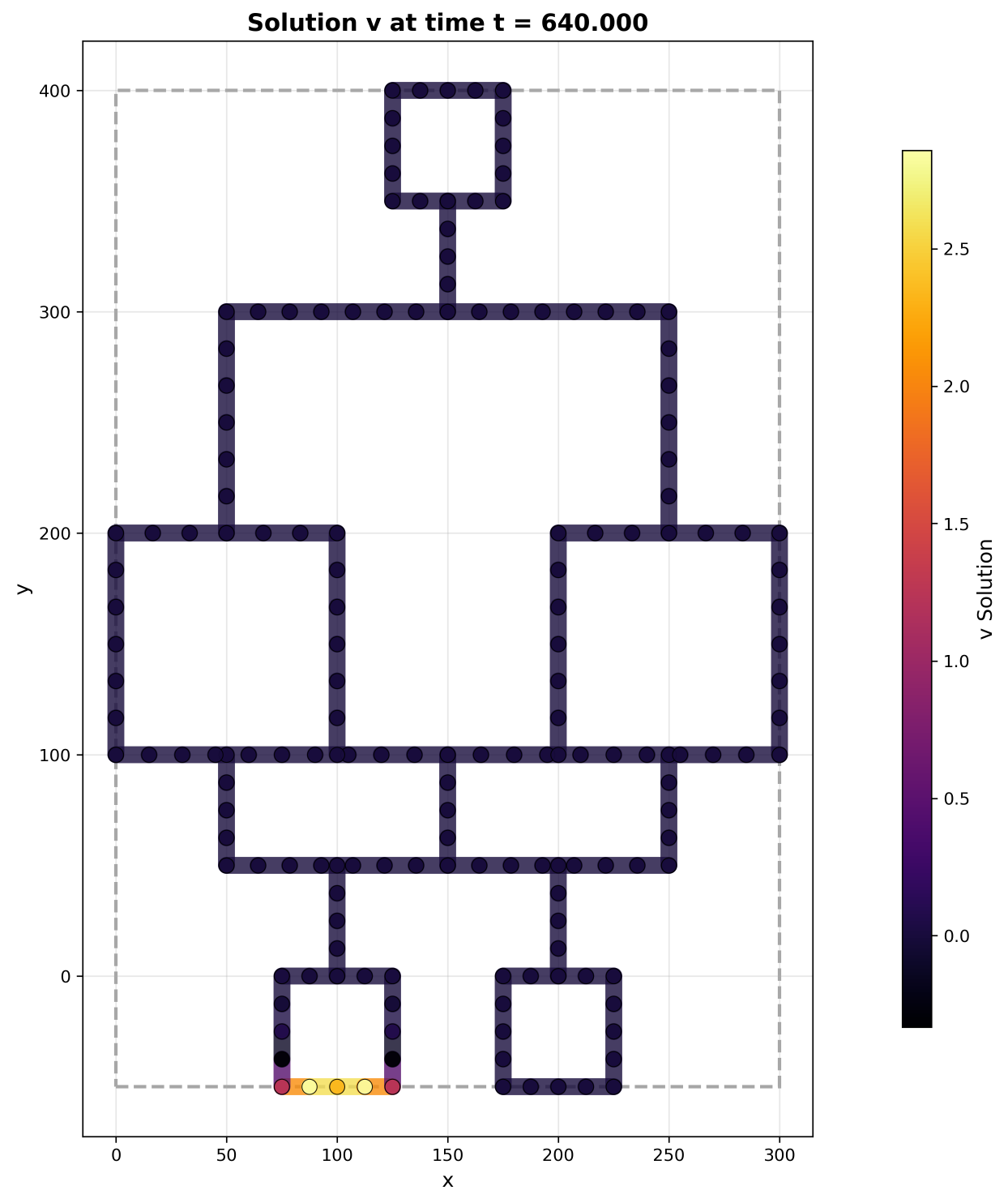}
    \caption{Tumour cells~$v$}
  \end{subfigure}\hfill
  \begin{subfigure}[t]{0.48\textwidth}
    \includegraphics[width=\textwidth]{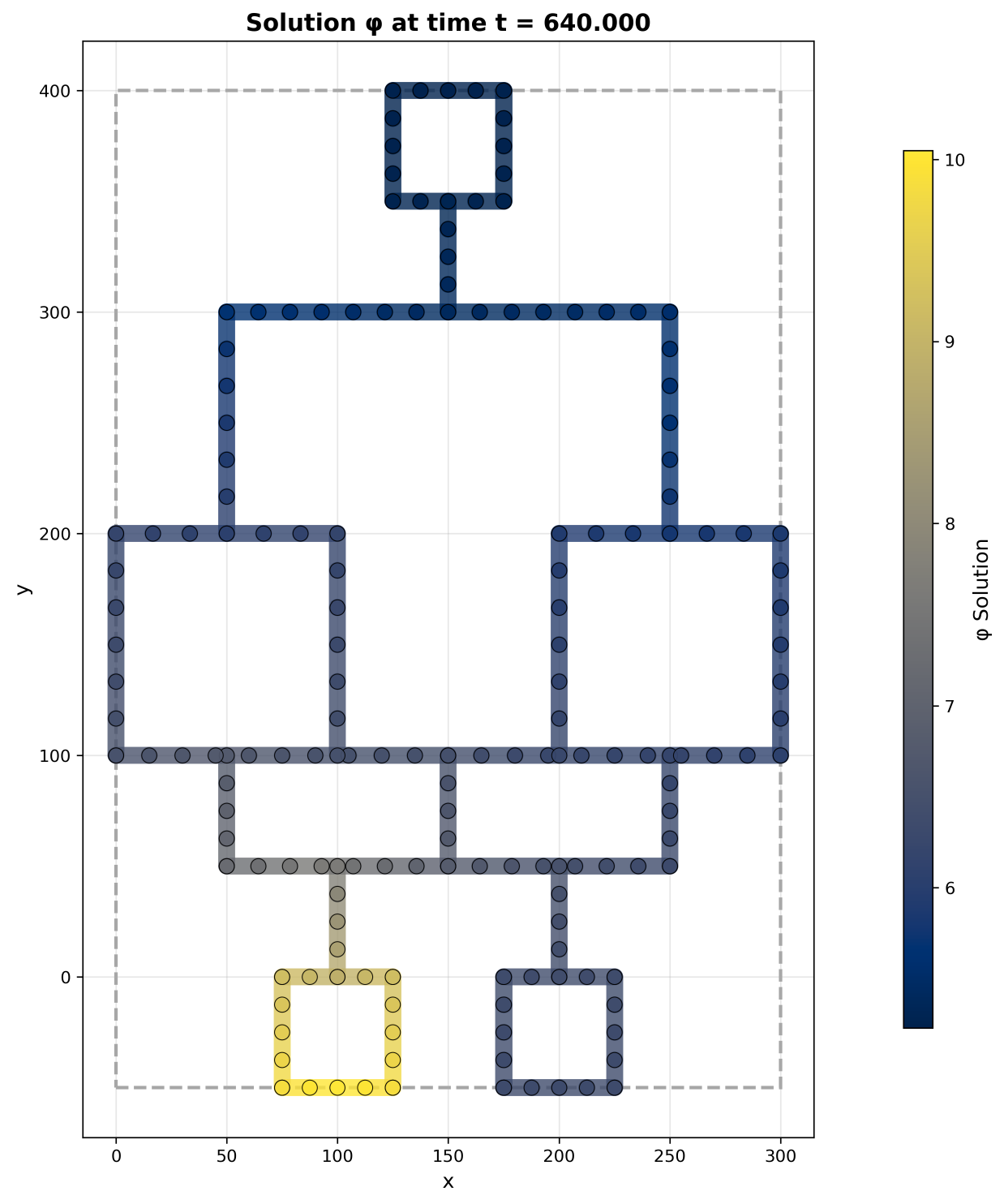}
    \caption{Chemoattractant~$\varphi$}
  \end{subfigure}
  \caption{Final state at $t=T=3600$ for all four unknowns.}
  \label{fig:final}
\end{figure}

Figure~\ref{fig:final} displays the bird's-eye view of all four
unknowns at the final time $t=T=3600$.

\subsection{Mass Conservation and Adaptive Time Stepping}

\begin{figure}
  \centering
  \begin{subfigure}[t]{0.48\textwidth}
    \includegraphics[width=\textwidth]{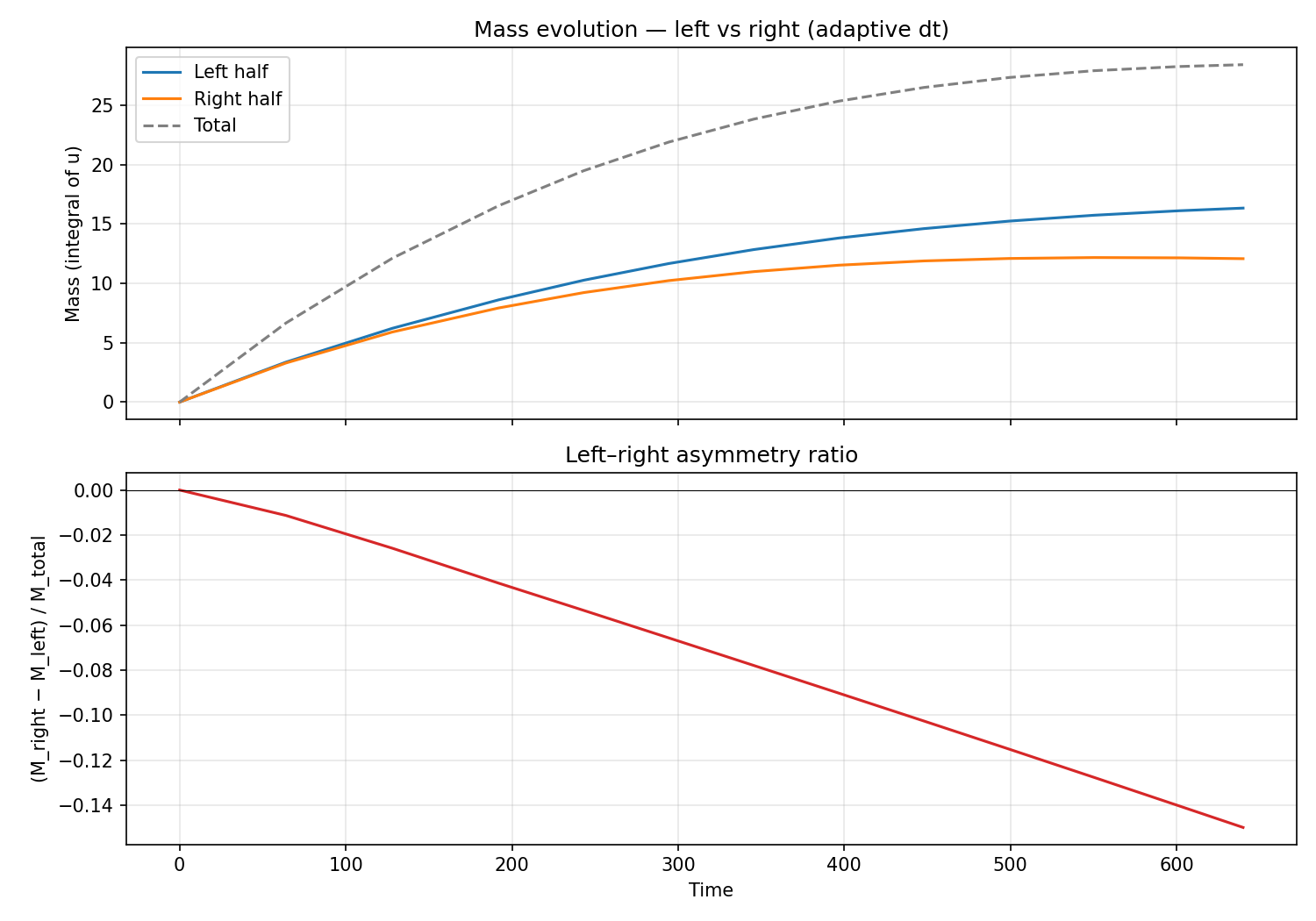}
    \caption{Left/right mass evolution of~$u$.}
  \end{subfigure}\hfill
  \begin{subfigure}[t]{0.48\textwidth}
    \includegraphics[width=\textwidth]{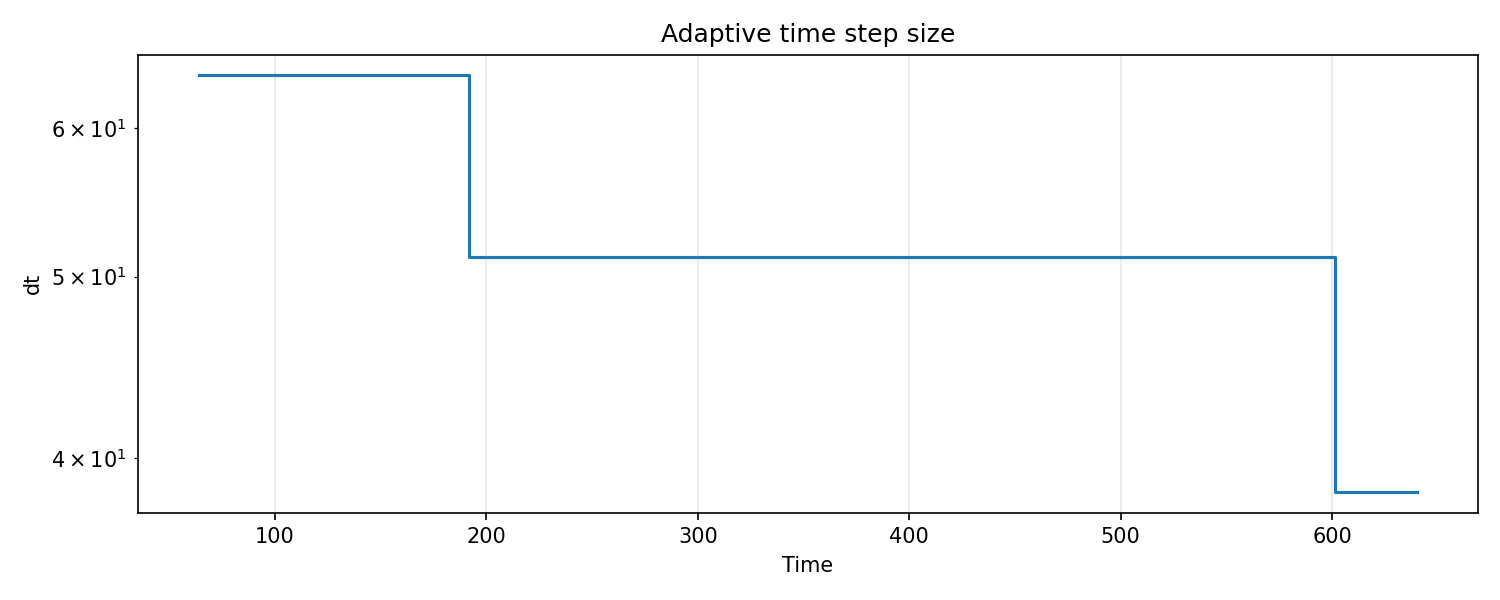}
    \caption{Adaptive time-step size~$\Delta t$.}
  \end{subfigure}
  \caption{Diagnostic plots for the OoC maze simulation with
           adaptive time stepping.  The mass plot verifies that
           the total immune-cell mass is redistributed between
           the left and right halves of the domain by diffusion
           and chemotaxis.  The $\Delta t$ plot shows how the
           adaptive controller increases the step size during
           smooth phases and reduces it when nonlinear effects
           require more Newton iterations.}
  \label{fig:diagnostics}
\end{figure}

Figure~\ref{fig:diagnostics} shows two diagnostic quantities:
the left/right mass evolution of the immune-cell density~$u$
(monitoring the integral of~$u$ over each half of the maze), and
the time-step size~$\Delta t$ as selected by the adaptive
controller.

\section{Getting Started}\label{sec:start}

\begin{lstlisting}[language=bash,caption={Installation}]
git clone https://github.com/silvia-bertoluzza/bionetflux
cd BioNetFlux
pip install -e .          # editable install
pip install -e ".[dev]"   # adds pytest, black, mypy
\end{lstlisting}

\begin{lstlisting}[language=Python,caption={Minimal Keller--Segel simulation}]
from bionetflux.setup_solver import quick_setup
from bionetflux.time_integration import TimeStepper
from bionetflux.visualization.lean_matplotlib_plotter import LeanMatplotlibPlotter

# 1.  Build the solver
setup = quick_setup("bionetflux.problems.ks_problem",
                    config_file="config/ks_parameters.toml")

# 2.  Initial conditions
traces, mults = setup.create_initial_conditions()

# 3.  Time integration
stepper = TimeStepper(setup)
for step in stepper.advance(traces, mults, n_steps=100):
    pass  # step.trace_solutions, step.time available here

# 4.  Visualise
plotter = LeanMatplotlibPlotter(setup.problems,
              setup.global_discretization.spatial_discretizations)
plotter.plot_birdview(step.trace_solutions, equation_idx=0,
                      time=step.time)
plotter.show_all()
\end{lstlisting}

\begin{lstlisting}[language=Python,caption={OoC maze with adaptive stepping}]
from bionetflux.setup_solver import quick_setup
from bionetflux.time_integration.time_stepper import AdaptiveTimeStepper
from bionetflux.geometry.domain_geometry import create_maze_geometry

geometry = create_maze_geometry(data_dir="maze_3_data", length=50.0)

setup = quick_setup("bionetflux.problems.ooc_problem",
                    config_file="config/ooc_maze3_parameters.toml",
                    geometry=geometry)

traces, mults = setup.create_initial_conditions()

stepper = AdaptiveTimeStepper(setup, dt_init=16.0, T_final=3600.0)
for step in stepper.advance(traces, mults):
    if step.step_number % 20 == 0:
        print(f"t={step.time:.1f}  dt={step.dt:.2f}  "
              f"Newton iters={step.newton_iterations}")
\end{lstlisting}

\section{Conclusions and Outlook}

\bionetflux{} provides a ready-to-use, extensible platform for
the simulation of reaction--diffusion--chemotaxis systems on
one-dimensional network geometries.  Its HDG discretisation yields
a compact global system (trace unknowns only) that scales well
with the number of domains, while the adaptive implicit time
stepper robustly handles the stiffness introduced by chemotactic
and reaction terms.

Current development directions include:
\begin{itemize}
  \item Higher-order ($p>1$) polynomial bases;
  \item Picard iterations for non linear static condensation;
  \item Hyperbolic models for CoC devices;
  \item Parallel element-local back-solves via multi-threading;
  \item Integration with experimental data pipelines for
        organ-on-chip calibration.
\end{itemize}

\section{Acknowledgements} 
In the preparation of this report, Claude (Claude 4.6 Opus. Anthropic) was utilized to analyze the provided software codebase and draft the initial summary of functionalities. The raw analysis was reviewed, refined, and verified for accuracy by the author. This work is realized with the support of the Italian Ministry of Research, under the complementary action NRRP “D34Health - Digital Driven Diagnostics, prognostics and therapeutics for sustainable Health care” (Grant \#PNC0000001). The author is member of the Istituto Nazionale di Alta Matemetica - Gruppo Nazionale di Calcolo Scientifico.

\begingroup
\raggedright

\endgroup

\end{document}